\documentclass[twocolumn]{aastex7}
\usepackage[yyyymmdd,hhmmss]{datetime}
\usepackage{CJK}
\usepackage{float}
\usepackage{booktabs} %For commands used in the tables
\usepackage{xargs} % Use more than one optional parameter in a new commands
\usepackage{chemformula}
\usepackage{graphicx}
\usepackage{xcolor, soul}
\usepackage{pifont}
\usepackage[colorinlistoftodos,prependcaption,textsize=tiny,textwidth=.5in]{todonotes}
\usepackage{mathtools}
\usepackage[T1]{fontenc}
\usepackage{tablefootnote}
\usepackage{hyperref}
\usepackage{xspace}

\newcommand{\A}{$\,\text{\r{A}}$}
\begin{document}

\title{JWST/NIRSpec Detects Warm CO Emission in the Terrestrial-Planet Zone of HD 131488}

\correspondingauthor{Cicero X. Lu}
\email{cicero.lu@noirlab.edu}

\author[0000-0001-9352-0248]{Cicero X. Lu}
\altaffiliation{Gemini Science Fellow}
\affiliation{Gemini Observatory/NSF’s NOIRLab, 670 N. A’ohoku Place, Hilo, HI 96720, USA}
\email{cicero.lu@noirlab.edu}

\author[0000-0002-4388-6417]{Isabel Rebollido}
\affiliation{European Space Agency (ESA), European Space Astronomy Centre (ESAC), Camino Bajo del Castillo s/n, 28692 Villanueva de la Ca\~{n}ada, Madrid, Spain}
\email{isabel.rebollidovazquez@esa.int}

\author[0000-0001-5638-1330]{Sean Brittain}
\affiliation{Department of Physics and Astronomy, 118 Kinard Laboratory, Clemson University, Clemson, SC 29634-0978}
\email{sbritt@clemson.edu}

\author[0000-0002-6881-0574]{Tracy Beck}
\affiliation{Space Telescope Science Institute, 3700 San Martin Drive, Baltimore, MD 21218, USA }
\email{tbeck@stsci.edu}

\author[0000-0002-8382-0447]{Christine H. Chen}
\affiliation{Space Telescope Science Institute, 3700 San Martin Drive, Baltimore, MD 21218, USA }
\affiliation{William H. Miller III Department of Physics and Astronomy, Johns Hopkins University, 3400 N. Charles Street, Baltimore, MD 21218, USA}
\email{cchen@stsci.edu}

\author[0000-0002-5885-5779]{Kadin Worthen}
\affiliation{William H. Miller III Department of Physics and Astronomy, Johns Hopkins University, 3400 N. Charles Street, Baltimore, MD 21218, USA}
\email{kworthe1@jhu.edu}

\author[0000-0002-5758-150X]{Joan Najita}
\affiliation{NSF’s NOIRLab, 950 N. Cherry Avenue, Tucson, AZ 85719, USA}
\email{joan.najita@noirlab.edu}

\author[0000-0002-6318-0104]{Chen Xie}
\affiliation{William H. Miller III Department of Physics and Astronomy, Johns Hopkins University, 3400 N. Charles Street, Baltimore, MD 21218, USA}
\email{cxie21@jh.edu}

\author[0000-0002-7050-0161]{Aoife Brennan}
\affiliation{School of Physics, Trinity College Dublin, the University of Dublin, College Green, Dublin 2, Ireland}
\email{brenna29@tcd.ie}

\author[0000-0001-9504-8426]{Amaya Moro-Martin}
\affiliation{Space Telescope Science Institute, 3700 San Martin Drive, Baltimore, MD 21218, USA }
\email{amaya@stsci.edu}

\author[0000-0002-1783-8817]{John Debes}
\affiliation{Space Telescope Science Institute, 3700 San Martin Drive, Baltimore, MD 21218, USA }
\email{debes@stsci.edu}

\author[0000-0002-1002-3674]{Kevin France}
\affiliation{Laboratory for Atmospheric and Space Physics, University of Colorado Boulder, Boulder, CO 80303, USA}
\email{kevin.france@colorado.edu}

\author[0000-0003-4705-3188]{Luca Matr\`a}
\affiliation{School of Physics, Trinity College Dublin, the University of Dublin, College Green, Dublin 2, Ireland}
\email{lmatra@tcd.ie}

\author[0000-0002-3191-8151]{Marshall Perrin}
\affiliation{Space Telescope Science Institute, 3700 San Martin Drive, Baltimore, MD 21218, USA }
\email{mperrin@stsci.edu}

\author[0000-0002-2989-3725]{Aki Roberge}
\affiliation{Astrophysics Division, Code 660, NASA Goddard Space Flight Center, Greenbelt, MD 20771, USA}
\email{aki.roberge-1@nasa.gov}

\shorttitle{HD 131488}
\shortauthors{Lu et al. 2025}

\begin{abstract}
We have obtained a high-resolution, JWST NIRSpec $2.87$ -- $5.14$ $\mu$m spectrum of the debris disk around HD 131488. We discover CO fundamental emission indicating the presence of warm fluorescent gas within $\sim10$ AU of the star. The large discrepancy in CO's vibrational and rotational temperature indicates that CO is out of thermal equilibrium and is excited with UV fluorescence. Our UV fluorescence model gives a best fit of $1150\,$K  with an effective temperature of $450$, $332$, and $125\,$K for the warm CO gas kinetic temperature within $0.5$, $1$, and $10\,$AU to the star and a gas vibrational temperature of $8800\,$K. 
The newly discovered warm CO gas population likely resides between sub-AU scales and $\sim\,10\,$AU, interior to the cold CO reservoir detected beyond $35\,$AU with HST STIS and ALMA.
The discovery of warm, fluorescent gas in a debris disk is the first such detection ever made. The detection of warm CO raises the possibility of unseen molecules (H$_2$O, H$_2$, etc) as collisional partners to excite the warm gas. We estimated a lower mass limit for CO of $1.25\times 10^{-7}\text{M}_{\oplus}$, which is $10^{-5}$ of the cold CO mass detected with ALMA and HST. We demonstrate that UV fluorescence emerges as a promising avenue for detecting tenuous gas at $10^{-7}$ Earth-mass level in debris disks with JWST.
\end{abstract}

\section{Introduction}

\noindent Gas in debris disks is now routinely detected through mostly interferometric observations with ALMA. Observations in the millimetric range show the presence of CO and C in $\sim$ 30 debris disks \citep{Moor+17,Rebollido+22}. This gas is expected to be cold and in most cases is located in the outer regions of the system. In contrast with this cold --mostly--  molecular, gas, a hot component close to the star is also observed in optical wavelengths, sometimes showing variations that could originate in exocomets \citep[e.g.][]{Hobbs+85, Welsh+18, Rebollido+20, Iglesias+18}. The two different populations of gas have been suggested to be connected by \cite{Rebollido+18}, raising the question of where is the warm gas in the intermediate regions. 

Ground-based observations in the near-IR targeting the CO ro-vibrational mode have detected absorption CO in $\beta$ Pic \citep{Troutman+11}. They detected absorption from v=0, which probes the same vibrational band as ALMA and found that the rotational levels were not thermalized beyond J=2 and did not detect any fluorescent CO. Therefore, \citet{Troutman+11} places a lower limit on the inner extent of the CO at $25$\,AU, suggesting an inner radius of the gas smaller than what was initially suggested from ALMA data \citep{Dent+14}. Other debris disk sources with ALMA detections \citep[e.g.][]{Moor2019, Marino2016, Rebollido+22} do not show CO in the near-IR or have never been investigated with high-resolution spectroscopy.

HD 131488 is an early A-type member of Upper Centaurus Lupus (UCL) within the larger Scorpius-Centaurus complex \citep{Preibisch+08} with a GAIA distance of $152.2$ pc. Kinematic analysis \citep{Zerjal+23} and isochrone fitting of F-type members \citep{Pecaut+12} suggest that the stars in this subgroup have ages $\sim$15$\pm$3 Myr. 

Detailed fitting of the HD 131488 infrared to millimeter Spectral Energy Distribution (SED) has revealed two circumstellar dust components, a hot component with a temperature $\sim$750 K in addition to the cooler $\sim$100 K component, typically found in debris disks \citep{Lisse+17, Melis+13}. The properties of the hot dust component were inferred from 2MASS $J$, $H$, and $K_s$ and \emph{WISE} Band 1 ($3.35$ $\mu$m) and 2 ($4.6$ $\mu$m) photometry and an IRTF SpeX $2.5$\,--\,$5$ $\mu$m spectrum. Those of the cold component were inferred from \emph{WISE} Band 3 (11.6 $\mu$m) and 4 (22.1 $\mu$m) and \emph{IRAS} 12, 25, and 60 $\mu$m photometry \citep{Melis+13}. Intriguingly, a Gemini T-ReCS $7.5$\,--\,$13$ $\mu$m spectrum from \citet{Melis+13} revealed an excess above the continuum emission model with a declining slope, suggesting the presence of a solid-state emission feature with a peak wavelength $<$7.5 $\mu$m. The carrier for the solid-state feature has not been definitively identified because the solid-state feature is not complete; however, the emission is consistent with carbonaceous grains. To date, only four debris disks of the $\sim$120 studied spectroscopically thus far have $6$\,--\,$10$ $\mu$m emission consistent with the presence of carbonaceous grains: HD 36546, HD 121191, HD 131488, and HD 148657 \citep{Lisse+17}. ALMA millimeter continuum observations indicated that the total dust mass is 0.32 $M_{\earth}$ \citep{Moor+17}.

The circumstellar dust in the HD 131488 disk has been spatially resolved both in scattered light using VLT SPHERE \citep{Pawellek+24,Xie+22} and thermal emission using ALMA. The ALMA millimeter continuum images have revealed an edge-on disk with a radius of $88$ $\pm$ $3$ AU and a width $46$ $\pm$ $12$ AU \citep{Moor+17}. The millimeter continuum images trace the location of the cold dust population that consists of large, millimeter-sized dust grains that are gravitationally bound to the star and remain near the "planetesimal birth ring" where they were created \citep{Hughes+18}. The SPHERE scattered light images have also revealed an edge-on disk however with a larger radius (110 $\pm$ 25 AU) although the exact location is more uncertain \citep{Pawellek+24}. Scattered light images trace the location of smaller, micron-sized dust grains that are more sensitive to radiation pressure and gas-grain interactions. 

ALMA observations have also revealed a large mass of CO, $0.089$\, $M_{\earth}$, in the HD 131488 disk corresponding to a large CO-to-dust ratio ($\sim$\,$0.3$, \cite{Moor+17}), substantially larger than the $\sim$0.01 expected in the interstellar medium. The measured $^{12}$CO to $^{13}$CO and $^{13}$CO to C$^{18}$O isotopologue ratios indicate that both the $^{12}$CO and $^{13}$CO are optically thick if the gas has interstellar isotope ratios (i.e. [$^{12}$C]/[$^{12}$C] = $77$ and [$^{16}$O]/[$^{18}$O] = $560$) \citep{Milam+05, Ayres+13}. Since the disk is nearly edge-on with an inclination $82$$\arcdeg$ \citep{Pawellek+24}, its circumstellar CO has also been studied using ultraviolet absorption line spectroscopy in the CO A-X bands using \emph{HST} STIS and COS. Detailed modeling of the CO absorption lines suggest the presence of a foreground slab of cold gas with an excitation temperature, $T_{ex}$ = $45$ $\pm$ $8$ K, and a kinetic temperature, $T_{rot}$ = $60$$^{+9}_{-10}$ K, consistent with absorption through the cold, outer disk \citep{Brennan+24}. \cite{Pawellek+24} show that the difference in the spatial distributions of the micron- and millimeter-sized dust grains can be explained if the smaller particles are radiatively driven to larger radii where they accumulate at the outer edge of the gas disk.

In addition to circumstellar CO, the ultraviolet absorption line studies have also revealed the presence of \ion{C}{1}, a photodissociation product of CO. The \ion{C}{1} and CO have a similar column densities along the line-of-sight, 17.4 cm$^{-2}$ compared with 18.0 cm$^{-2}$ \citep{Brennan+24}. Since the measured \ion{C}{1} to CO ratio is lower than expected from a steady-state, collisionally generated CO, \cite{Brennan+24} have suggested that either the \ion{C}{1} is not well mixed throughout the midplane, or there is an additional \ion{C}{1} removal process. One possibility that has been discussed for the CO observed in young debris disks around A-type stars is that the CO is left over from the gas-rich protoplanetary disk phase \citep{Nakatani+23}. HD 131488 has a $L_{IR}/L_*$ (= 5.5$\times$10$^{-3}$, \cite{Moor+17}) which is in between the typical range for debris disks (10$^{-5}$ -- 10$^{-3}$) and protoplanetary disks (typically $\geq$0.1). For context, \emph{Spitzer} MIPS 24 and 70 $\mu$m photometry indicates that the UCL A-type star disk fraction is $\sim$25\% and that their typical fractional infrared luminosities are ($L_{IR}/L_*$ $<$ 10$^{-5}$-10$^{-3}$) \citep{Chen+12}. This disk fraction is slightly higher than the field debris disk fraction ($\sim$10\%), consistent with the young age of UCL \citep{Wyatt+08}, and this range of $L_{IR}/L_*$ is consistent with that typically measured for debris disks. 

\begin{figure*}[ht!]
    \centering
    \includegraphics[scale=0.52]{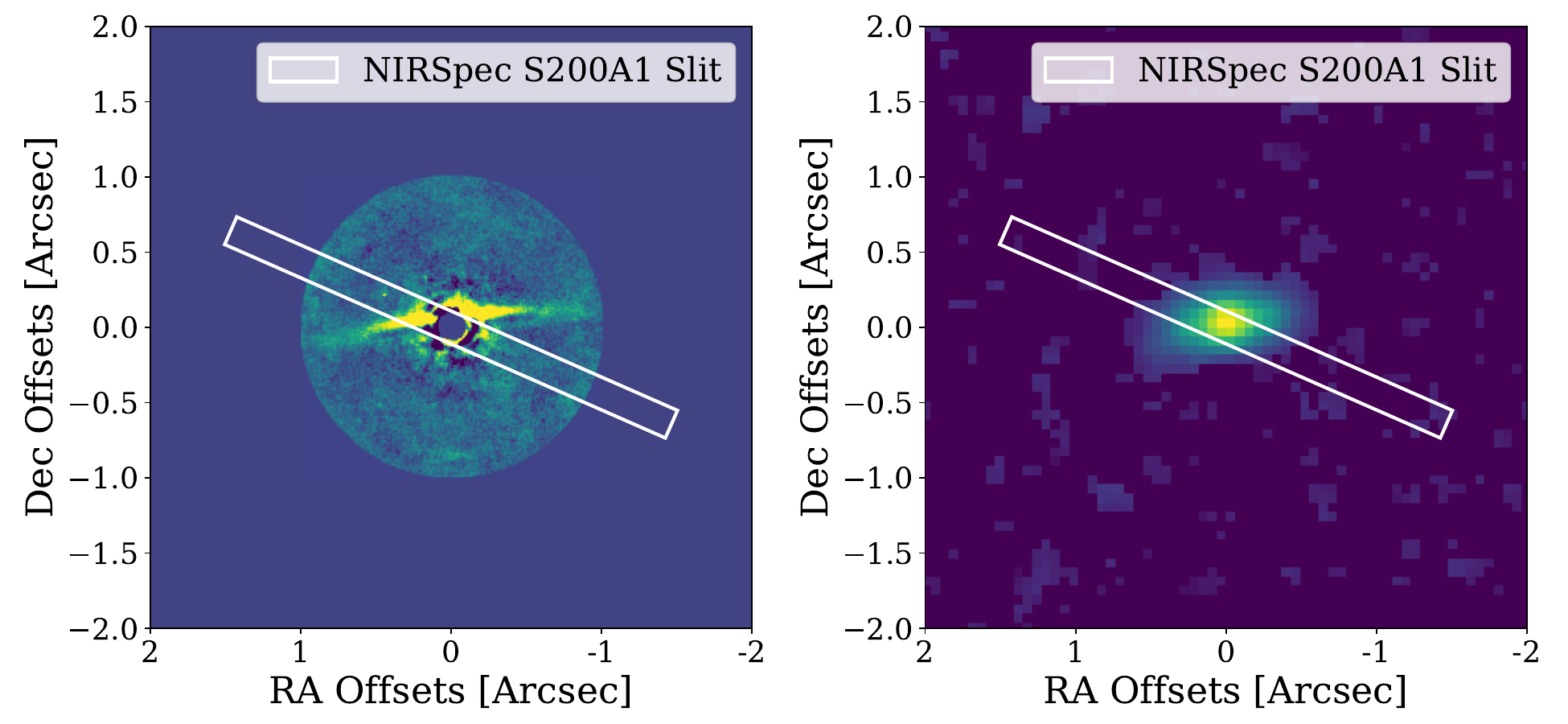}
    \caption{HD 131488 Disk and Slit Alignment, showcasing our data is sensitive to the disk region inward of 30 AU to the star. \textbf{Left}: The JWST NIRSpec Slit is overlaid on VLT/SPHERE coronagraphic imaging of scattered light of dust in the HD 131488 disk. The image has been post-processed with Non-negative matrix transformation \citep{Xie+22}.  The alignment between the nearly edge-on disk and the PA of the slit clearly shows that our data is sensitive to the inner region of the disk. \textbf{Right}: Overlay of NIRSpec slit on ALMA $^{12}$CO(2-1) image of HD 131488 \citep{Moor+17}. 
    The major and minor axes of the disk are assumed to be $1.14^{\prime\prime}$ by $0.7^{\prime\prime}$ according to the disk profile reported in the ALMA CO image \citep{Moor+17}. For both gas disk and dust disk, our data is sensitive to the disk region inward of a radius of $15$ AU to the star, because the star and inner region of the disk fall in the slit, but are insensitive to the outer region of the disk, because those regions fall outside of the slit.}
    \label{fig:slit-alignment}
\end{figure*}

% \begin{figure*}[ht!]
%     \centering
%     \includegraphics[scale=0.52]{firstlook.pdf}
%     \caption{Figures with CO bands highlighted. 2 panels, top: continuum subtracted}
%     \label{fig:first-look}
% \end{figure*}

Our collaboration is searching for CO fundamental absorption or emission in debris disks with ALMA detected CO. The JWST GO 2053 program (PI Rebollido) is using the NIRSpec Fixed Slit to observe HD 36546, HD 110058, HD 131488, HD 131835, and HD 156623 -- five edge-on debris disks with time-variable metal lines associated with the presence of exocomets \citep{Moor+17,Rebollido+22}. JWST GTO program 2780 (PI Gaspar using GTO time from MIRI US Investigator Chen) is using the NIRSpec Fixed Slit to observe another edge-on debris disk with time-variable metal lines, HD 32297. JWST GO program 1563 (PI Chen) is using the NIRSpec IFU to observe 49 Cet, $\beta$ Pic, and HD 181327. This manuscript is the first in a series describing the results from the search for CO and the detailed characterization of the near-infrared continuum. In Section \ref{sec2}, we describe the \emph{JWST} observations and data reduction. In Section \ref{sec3}, we present CO modeling of the ro-vibrational lines detected in the \emph{JWST} spectrum, and present our results in Section \ref{sec4}. In Section \ref{sec5}, we discuss how our measurement impacts our understanding of the origin of the gas and also discuss it in the context of debris disks' predecessors, the protoplanetary disks. In Section \ref{sec6}, we present our conclusions.

\section{Observations and data reduction}\label{sec2}
We observed HD 131488 ($K_{s}$ = 7.803) on 11 February 2023 as part of JWST Cycle 1 General Observer Program 2053 (PI: Rebollido). The goal of the program was to search for absorptions of CO ro-vibrational modes around 4.5 $\mu$m in a set of edge-on debris disks with previous ALMA CO detections \citep{Rebollido+22}. Observations were obtained with NIRSpec Fixed Slit, using the S200A1 slit (0.2$\arcsec$$\times$3.3$\arcsec$) and the G395H/F290LP disperser/filter combination (2.87-5.14 $\mu$m; R$\sim$2,700). We used Wide Aperture Target Acquisition (WATA) on the target to precisely place the star in the center of the slit to ensure accurate wavelength calibration. Since NIRSpec is spectrally undersampled over a large wavelength range, we dithered the source in the spatial and spectral directions using 3 primary nod positions in the spatial direction and 3 subpixel positions in the spectral direction for a total of 9 exposures, in order to provide improved subpixel sampling. Finally, we used the SUBS200A1 subarray to prevent saturation of the bright target star.

Our Fixed Slit (FS) spectrum of HD 131488 is sensitive to inner regions of the disk, which allows us to place constraints on the geometry of the emitting regions for modeling. Figure \ref{fig:slit-alignment} shows the alignment between the disk and slit position angles (PA). Fig. \ref{fig:slit-alignment} left panel shows JWST/NIRSpec slit overlaid on VLT/SPHERE coronagraphic imaging of scattered light of dust in the HD 131488 disk \citep{Xie+22}. Similarly, Fig. \ref{fig:slit-alignment} right panel shows the slit overlaid on the ALMA gas disk of $^{12}$CO(2-1) emission from \citet{Moor+17}. The calculation of the misalignment between disk and slit PA empirically constrains the emission radii to within $30\,$AU of the disk. We examined the 2D spectrum at the detector level and found that the CO lines are not spatially resolved compared to NIRSpec FS standard stars from the commission program (PID: 1128), which empirically limit the inner emission radius to be within $\leq\,15\,$AU, a resolution element for the beam of the JWST/NIRSpec FS observation. In section \ref{sub:geometry}, we further expand on the geometry constraints of the model empirically measured from data.

\subsection{Data reduction}

The JWST NIRSpec Fixed Slit (FS) mode delivers single object spectroscopy with the highest sensitivity achievable with NIRSpec.  For our data processing, the "Stage 2" pipeline was run on each integration of the HD 131488 spectral files using the 1.12.3.dev21+g2199ba88 version.  This was the first development version of the JWST calibration pipeline that included updated and proper reference files for slit loss corrections for NIRSpec FS data processing.  For optimal signal-to-noise on HD 131488, we rewrote the further NIRSpec FS "Stage 3" processing methodology.  In the default reduction pipeline, the curved and tilted FS spectra are resampled onto a regular, linear pixel grid for every exposure before being combined and extracted into 1-dimensional (1D) data products.   We instead used a cross-dispersed profile weighted 1D extraction and combination of 1D spectra merged onto a common pixel grid.  For bright point sources, this "extract to 1D and then combine" strategy can vastly improve the signal-to-noise (S/N) over the "resample in 2D, combine and then extract" method used in the default NIRSpec pipeline.

For our processing, we created a cross-dispersion point spread function (PSF) shape from un-rectified "Stage 2" pipeline products ("cal.fits" files). We created a PSF for each spectral column in the data using a sigma clipped mean, scaled this profile to the peak of the spectral trace, and summed a 5 pixel aperture across the spatial profile in each spectral pixel.  This trimmed out flux outliers and improved the 1D extracted spectral product compared to a simple pixel sum.  The result of this extraction was a 1D flux and non-linear wavelength array for each of the 2 integrations in the 9 dither exposure pattern, for a total of 18 unique measures of the 1D spectrum of HD 131488.  

Combination of the individual 1-D non-linear spectral exposures into a single linear product was accomplished by merging the 18 exposure-level arrays into a single array of 18 times the size, sorted by wavelength.  The final spectrum was created by taking the mean flux rate in a grid of linearly sampled wavelength elements; no spectral oversampling was used. We identified and clipped out significant flux outliers from hot pixels and detector effects that were common in primary dither positions.  The standard deviation of the flux values for each linear spectral element was used to define the noise per spectral pixel.  In the 4.0-5.0$\mu$m wavelength region, the final combined data product had an S/N of $\sim$ 150.
% 300.

\section{Analysis}\label{sec3}
\begin{figure*}[ht!]
\epsscale{1.2}
    \plotone{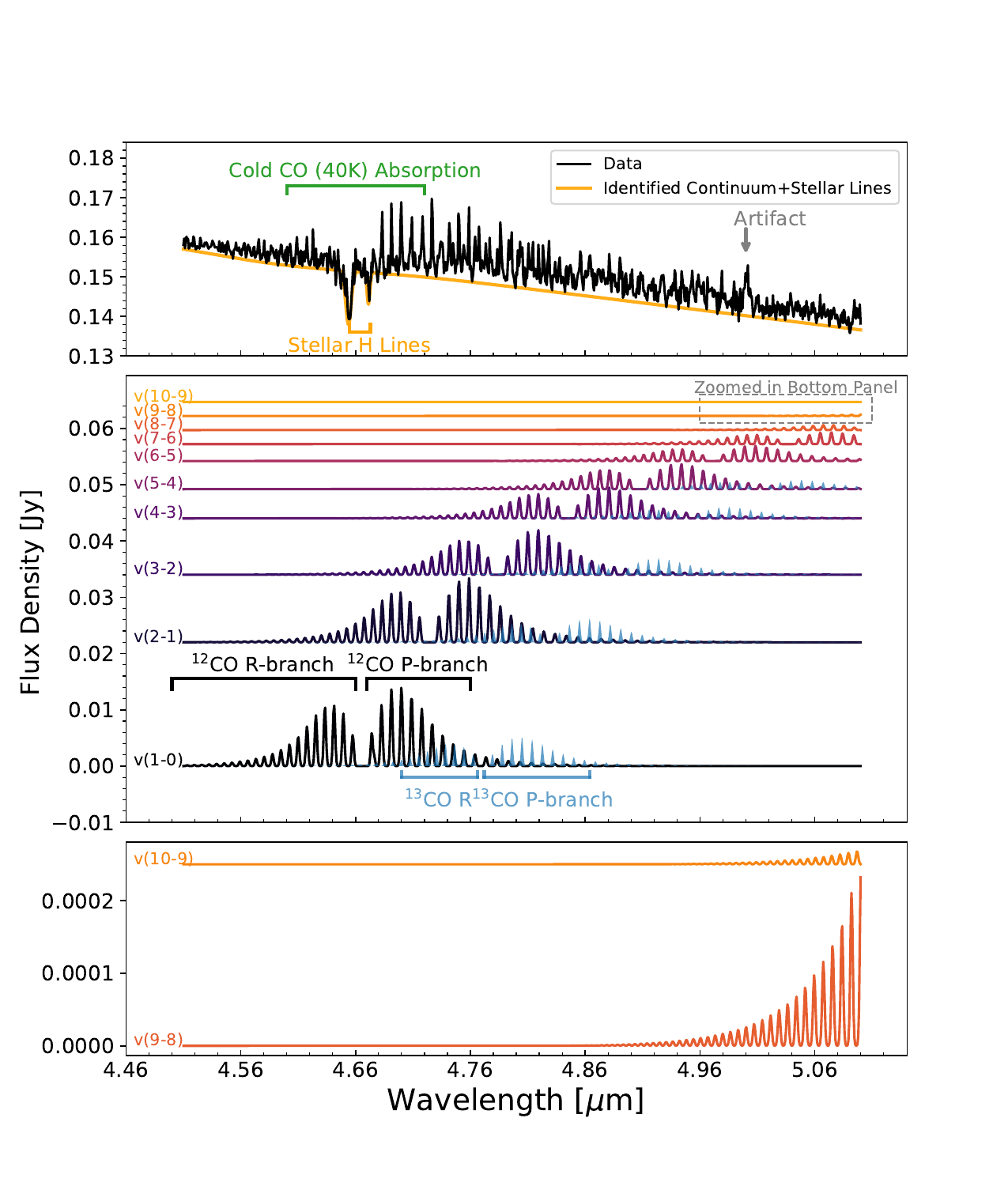}
    \caption{A first look at HD 131488 NIRSpec spectrum reveals the presence of multiple vibrational levels of CO ro-vibrational emission, indicating UV fluorescence. \textbf{Top}: HD 131488 NIRSpec spectrum plotted against an identified stellar continuum. Regions affected by cold CO absorption are annotated with a green bracket and text. Features such as artifacts and stellar hydrogen lines are also annotated. \textbf{Middle}: Models of individual vibrational levels are vertically offset for clarity, with $^{13}$CO contributions shaded in opaque blue.  \textbf{Bottom}: Zoomed in view for v(9-8) and v(10-9) models. \added{The individual lines or features detected at 5-$\sigma$ after stellar photospheric subtraction are shown in Figure \ref{fig:sigmaClip} and detailed in section \ref{subsec:5sigma}.}  }
    \label{fig:vib-Levels}
\end{figure*}

\begin{figure*}[th!]
    \centering
    \includegraphics[scale=0.78]{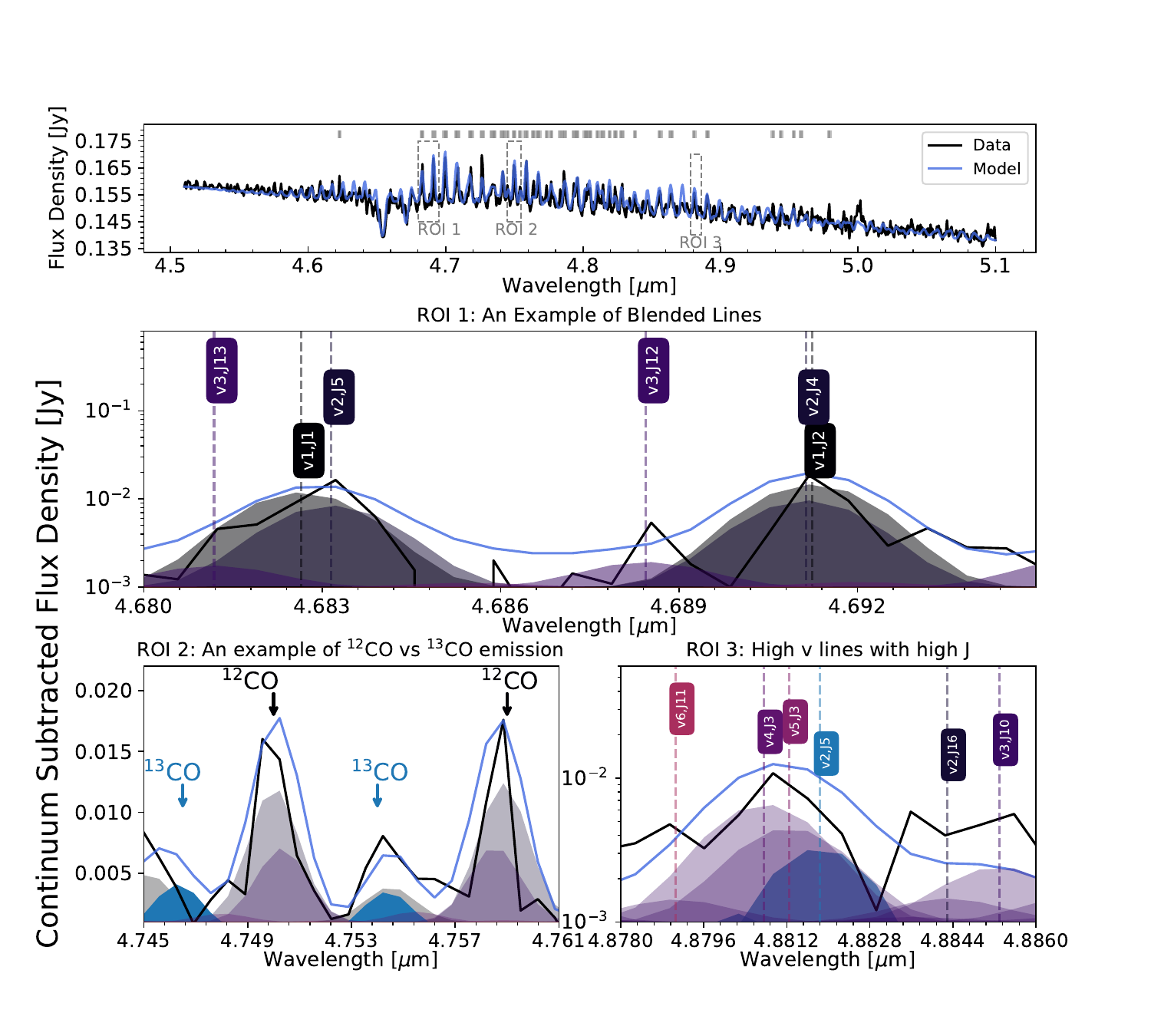}
    \caption{\textbf{HD 131488 CO model (blue) compared with NIRSpec data (black).} \textbf{The top panel} shows the same best-fit model as in Figure \ref{fig:COmodel}, while the remaining panels highlight zoomed-in regions of the spectrum. The short vertical gray lines at 0.176 Jy indicate line or features detected at $5\sigma$ over disk continuum after stellar photosphere subtraction. \textbf{Middle}: Example of a blended line (ROI 1), where each observed feature contains contributions from multiple vibrationally excited transitions. \textbf{Bottom left}: Example of $^{12}$CO versus $^{13}$CO emission (ROI 2), with $^{13}$CO lines shaded in blue and $^{12}$CO vibrational levels shaded according to the scheme in Fig. \ref{fig:vib-Levels}. \textbf{Bottom right}: Example of \added{moderate to high} rotational lines at highly excited vibrational levels (ROI 3) that constitute features detected at $5~\sigma$ in the observed spectrum.}
    \label{fig:examples}
\end{figure*}

In this section, we first describe the observed CO spectral features qualitatively, infer the excitation mechanisms, and diagnose the line blending issues from the line flux measurements. We then proceed to model CO features with a two-component model,  a UV-fluoresced warm gas component and a foreground cold gas absorption component. We also discuss the optical depth of gas and dust. We rule out the effects of foreground ISM absorption and extinction on the observed spectrum.  

\subsection{A High-level Overview of the Spectrum}
We provide an intuitive, high-level explanation of the spectra and annotate the prominent spectral features before diving into the details of our modeling. Through Figure \ref{fig:vib-Levels} and \ref{fig:examples}, we highlight four important observations that will be further elaborated and explored in the following subsections in this analysis section.

Firstly, we detect multiple vibrational levels of $^{12}$CO and $^{13}$CO, confirming UV fluorescence. Figure~\ref{fig:vib-Levels} (middle and bottom) shows $10$ levels of $^{12}$CO from v(1–0) to v(10–9) and $5$ levels of $^{13}$CO. Each is annotated and vertically offset for clarity; relative strengths are plotted to scale, with solid curves for $^{12}$CO and blue shading for $^{13}$CO.  We also annotate the locations of the artifact and the P- and R-branch of CO emission using the v(1-0) transition. 
It is worth noting that the prominent emission line at $5\,\mu$m is likely a detector artifact. Similar $5\,\mu$m lines exist in all spectra in JWST PID 2053 (PI: Rebollido), but at least one source's MIRI spectrum (GTO program, PI: Henning) at commensurate wavelength does not show such a feature. In Figure~\ref{fig:examples} (top), we highlight three regions of interest (ROI) to show the effect of line-blending, the detection of istopologues, and the presence of UV-fluoresced warm gas. ROI~1 demonstrates line blending: a single feature may arise from multiple vibrational and rotational transitions. For example, the leftmost feature (middle panel) blends three lines, v(1–0) J(1–0), v(2–1) J(5–4), and v(3–2) J(13–12). Similarly, the $4.6911\,\mu$m feature results from several vibrationally excited transitions near the same wavelength. A quantitative discussion of line blending appears in subsection~\ref{sec:line-flux}.

Secondly, we detect $^{13}$CO alongside $^{12}$CO, as $^{12}$CO alone cannot reproduce the observed line ratios. ROI~2 in Figure~\ref{fig:examples} (bottom left) illustrates the relative contributions of both species. $^{13}$CO lines are shaded in blue regardless of vibrational level, while $^{12}$CO lines follow the vibrational color scheme in Figure~\ref{fig:vib-Levels}. Weighting all lines by their SNR, we fit for the $^{12}$CO/$^{13}$CO isotopologue ratio. The modeling approach is described in Section~\ref{sub:CO-model}, with implications discussed in Section~\ref{sub:gasimplications}.

In addition, we find evidence for CO gas at both warm and cold temperatures. Highly excited rotational transitions, such as v(7–6) J(17–16), v(4–3) J(5–4), and v(6–5) J(7–6), are annotated in Figure~\ref{fig:examples} (bottom right) and demonstrate the presence of warm gas, while ground-state transitions trace colder gas (lines in ROI~1). To reconcile these components, we investigate a temperature gradient and disk geometry in Section~\ref{sub:geometry}. We show the resulting model spectra from various regions of the disk in Figure~\ref{fig:COmodel}. 

Finally, we observe possible CO absorption, as suggested by asymmetric band heads and the annotated region affected by absorption in Figure \ref{fig:vib-Levels} top panel. A simple summation of modeled fluxes in Figure~\ref{fig:vib-Levels} middle panel overpredicts the line flux between $4.56$ and $4.7\,\mu$m. Following the cold CO absorption detected with \textit{HST}/STIS by \citet{Brennan+24}, we explore whether a foreground cold CO component may absorb flux in the $^{12}$CO fundamental R-branch. Constraints on its temperature and column density from STIS and past ALMA observations are applied in Section~\ref{sub:CO-model}, and a schematic representation is provided in Figure~\ref{fig:cartoon}.

\subsection{UV Fluorescence Emission of CO}
Prominent ro-vibration CO emission is observed spanning $4.6$ -- $5.1$\micron. The forest of emission lines indicates a population of multiple vibrational levels (as shown in Fig. \ref{fig:vib-Levels} middle and bottom panels), and the absence of high-J v=1-0 R-branch CO lines indicates that the rotational temperature is significantly lower than the vibrational temperature. This signature of UV fluorescence has been observed in Herbig stars with transition disks such as HD~141569 \cite[e.g.][]{Brittain+03, jensen2021} and HD~100546 \cite[e.g.][]{Brittain+09}, among others. In these cases, the UV continuum near 1500\AA\ excites the CO gas into electronically excited bound states. As the molecules relax to the ground electronic state, the vibrational levels are populated in such a way to reflect the diluted color temperature of the UV radiation field \citep{krotkov1980}. The rotational population of the molecules reflects the rotational population of the ground state because the selection rule for transitions between rotational levels, $\Delta J=\pm1$, applies and the rotational transitions are slow relative to the vibrational and electronic transitions. 

Traditionally, measurements of unblended CO line flux are used to infer the rotational and vibrational temperatures of the gas \citep{Brittain+03, jensen2021}. 
In the following section, we quantitatively analyze the observed emission line profile to assess whether a line-based approach can be applied to our NIRSpec data.

\subsection{Diagnosing Line Blending Issue from Line Flux Measurements}\label{sec:line-flux}
% NIRSpec G395H/F290LP disperser/filter wavelength range covers the entire CO fundamental, R and P ro-vibrational branches, allowing us to identify the lines and to infer the upper state quantum number of the transitions. 

In this subsection, we evaluate the impact of line blending at NIRSpec's spectral resolution and explain our rationale for deciding against a line-based method in constructing the rotational diagram.
Most CO lines typically become spectrally un-blended at R$\,\sim\, 20,000$, for a typical line velocity of 15 km/s for debris disks at 4.6$\mu$m. Working with our R$\sim\,2700$ spectrum \citep{Jakobsen+2022} with a FWHM of $0.00145$\,\micron, 
we measure the line fluxes for CO lines that appear unblended at first sight at the NIRSpec G395H resolution. We identify observed line positions and compare them with the rest wavelength line positions. We consider the observed line to be blended if the line position is separated from any adjacent lines in the v(3-2), v(2-1), and v(1-0) transitions by at least one full width at half maximum (FWHM). Given the spectral resolution, we assume the FWHM of absorption line profiles to be $1.45 \times 10^{-3}\,\mu$m, which corresponds to $\sim100\,\mathrm{km}\cdot s^{-1}$. If a line passes the blending criteria, we then fit a Gaussian to the line profile and calculate the flux using the equation
\begin{equation}
    F(v^{\prime\prime}\rightarrow v^{\prime}, J^{\prime\prime}\rightarrow J^{\prime}) = \int_{\lambda_{obs}-\sigma}^{\lambda_{obs}+\sigma} A\,\mathrm{exp}\frac{(x-\lambda_{obs})^2}{2\sigma^2},
    \label{eq:line-flux}
\end{equation}
where $\sigma$ is the root mean square (RMS) Gaussian width of the line, and the full-width at half maximum (FWHM) for thermal broadening is equal to $2.35$\,$\sigma$. $\lambda_{obs}$ is the observed line position, and A is the observed line amplitude from the continuum baseline. The main source of uncertainty in line flux measurements is the uncertainties present in the baseline continuum flux. Due to severe line-blending issues, we calculate the line flux uncertainties by taking into account the average standard deviations in parts of the spectrum on the NRS2 detector where we do not expect any molecular or atomic features. We measure the SNR to be $150$ and assigned a $0.67\%$ error to $\sigma$ in Equation \ref{eq:line-flux}.

We find that none of the emission lines in our NIRSpec spectrum is spectrally resolved. We tabulate the line ID and measured fluxes in Table \ref{tbl:line-flux} in the Appendix. When we assume only $^{12}\text{CO}$ (3-2), (2-1), and (1-0) transitions in the spectrum, we can identify 7 unblended lines for the R branch and 5 unblended lines for the P branch of the $^{12}\mathrm{CO}\,v(1-0)$ transitions. But when we take into account the $^{13}\text{CO}$ (3-2), (2-1) and (1-0) transitions, there are only 2 and 3 unblended lines for R and P branches, respectively. However, when we include the CO population for $0\leq J \leq 120$ and $0\leq v \leq 10$, we find that even lines previously identified as unblended appear to be blended. We show an example in Fig. \ref{fig:model_annotated} in the Appendix. 

The severe line blending issue prevents us from estimating the rotational and vibrational excitation temperatures of gas from individual line fluxes. In the subsequent section, we turn away from the line identification and instead use an alternative, phenomenological modeling approach that takes into account  $^{12}\mathrm{CO}$ and $^{13}\mathrm{CO}$ transitions and model the CO emission region from $4.5$--$5.1\,\mu$m simultaneously. 

\subsection{CO Fluorescence Modeling}\label{sub:CO-model}
\begin{figure*}[ht!]
\epsscale{1.2}
    \plotone{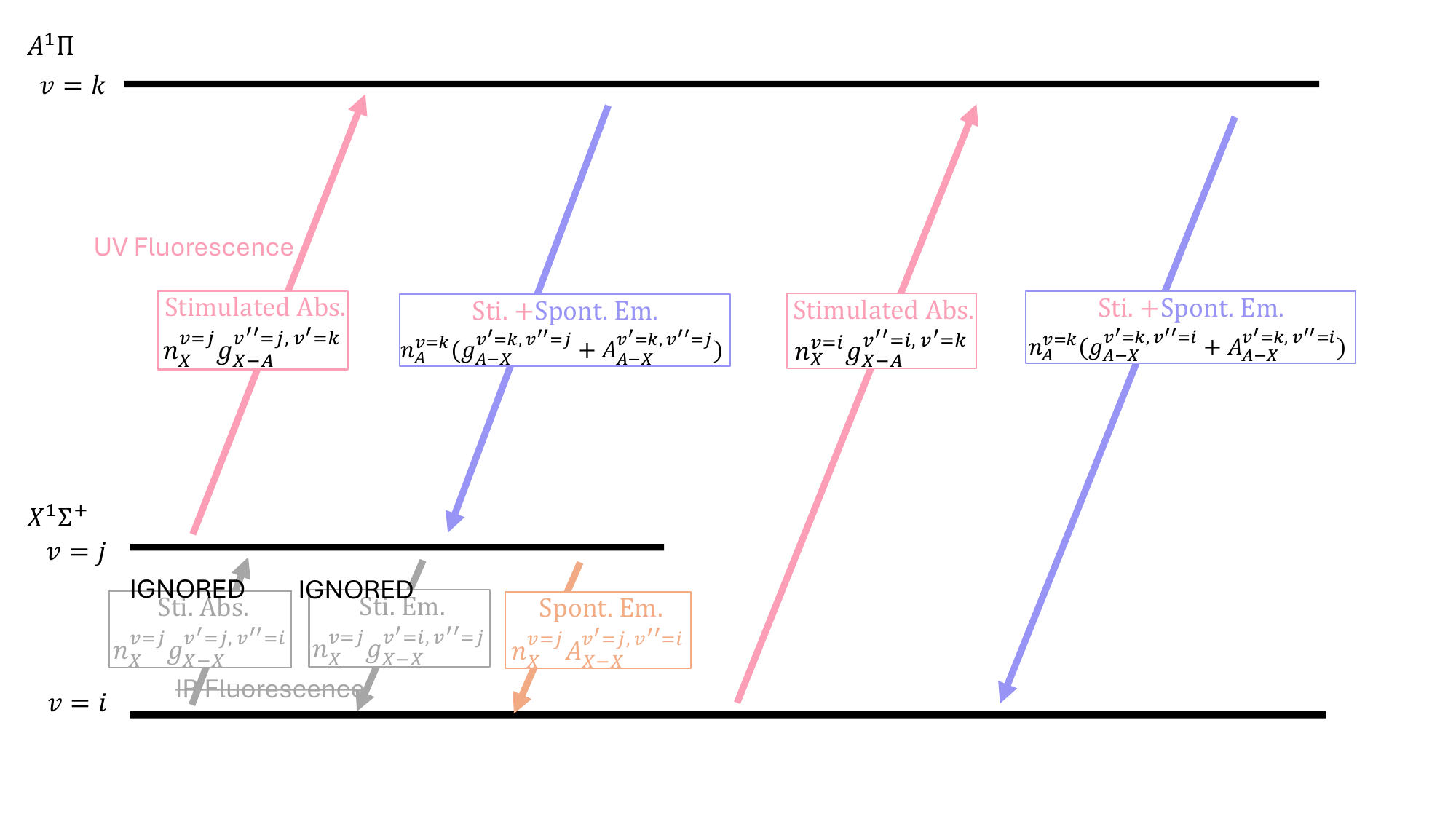}
    \caption{Energy Level Schematic Diagram of CO. In this diagram, we consider stimulated absorptions and stimulated+spontaneous emissions driven by UV fluorescence. We choose to ignore IR fluorescence because its transition probability is a few orders of magnitude smaller than that of UV fluorescence. $n_{X}$ and $n_{A}$ stands for the level population of CO at the ground and first excited electronic state $X^{1}\Sigma^{+}$ and $A^{1}\Pi$. $A_{A-X}$ stands for Einstein A coefficients for spontaneous emission (in purple), and $g_{A-X}$ and $g_{A-X}$ stands for Einstein B coefficients for stimulated emission and absorption (in pink). For application to HD 131488, we only consider fundamental transitions which means $k=j+1$.}
    \label{fig:schematic}
\end{figure*}
In section \ref{sec:line-flux}, we show that the spectrum of the ro-vibrational CO emission from HD~131488 is spectrally unresolved, so the radial distribution of the gas cannot be inferred from the line profile. Moreover, because of the extensive line blending, it is impossible to directly measure the rotational and vibrational temperatures of the gas from the individual line fluxes. Thus, we simultaneously model all CO emission and absorption lines in the spectrum using a high spectral resolution and down-sample it to the NIRSpec resolution. 

The model produces a spectrum by adopting a slab of gas where the rotational populations are described by,

\begin{equation}
    N_J=N_v g_J' e^{\frac{-E^{\prime}_J}{kT_{rot}(r, \alpha)}}
\end{equation}
where $N_J$ is the column density of molecules in state $J$ within a vibrational band, $g_J$ is the statistical weight of the level given by $2J+1$, E$^{\prime}_J$ is the rotational energy with the vibrational band, and $T_{rot}$ is the rotational temperature. $N_v$ is the column density of the CO molecules in a vibrational level given by,

\begin{equation}
    N_v=N\frac{e^{\frac{-E^{\prime}_v}{kT_{vib}(r, \beta)}}}{Q} \,n_{v}, 
\end{equation}

where $N$ is the total column density of CO, $E^{\prime}_v$ is the energy 
of the vibrational level, $T_{vib}$ is the vibrational temperature, and $Q$ 
is the partition function and $n_{v}$ is the fractional level population. The populations are calculated for $0\leq J \leq 120$ and $0\leq v \leq 10$. The partition function is calculated by ensuring that the sum over all states equals $N$. $n_{v}$ is calculated using equations \ref{eq:oscillator}, \ref{eq:transition_prob} and \ref{eq:groundE} , described in the paragraphs below. 

The emission lines are assumed to have a Gaussian profile with an intrinsic line width of $\sim\,1$~km~s$^{-1}$ as measured from ALMA J = (2\,--\,1) $^{12}$CO emission in \citet{Moor+17} and to be optically thin. Thus the flux of the emission scales as 
\begin{equation}
    F^{v,j}= \frac{h\, \nu\,A_{v,j}\,N_v\,\Omega}{4\pi},
\end{equation}
where \added{$\nu$ is the frequency} of the transition, $A_{v,j}$ is the Einstein A coefficient, and $N_v$ is the column density of CO molecules in state $v$, and $\Omega$ is the surface area of the emitting region. The resulting flux per v, j transition is $F^{\nu, j}$ is in units of \added{erg$\cdot s^{-1} \cdot cm^{-2}$}. 
\added{We then convert the integrated line flux to flux density, which has the units of erg$\cdot s^{-1} \cdot cm^{-2}\cdot \mu\textrm{m}^{-1}$, to match the JWST spectrum, since it has units of flux density.}

We find that using a single slab of gas with a single rotational and vibrational temperature fails to reproduce the relative intensities of the lines and highly asymmetric R and P branch line fluxes in the spectrum. We add an absorption slab component described by 2 parameters -- an absorption column density (N$_{\rm absorption}$) and a rotational temperature (T$_{abs}$) to the model, motivated by ALMA cold gas detections \citep{Moor+17, Brennan+24} in both millimeter and the UV wavelengths. The multi-wavelength observations reported a consistent column density of $\log \bar{N}_{\rm abs}(^{12}CO)=18.1^{+0.2}_{-0.1}\, cm^{-2}$ and kinetic temperature of $(45\pm\,9\,)\,$K. Since the ground state levels will also absorb infrared emission, we add foreground gas at this temperature and assume it absorbs the emission we see. We calculate the absorption spectrum with optical depth $\tau$
\begin{equation}
    \tau = \frac{g_u\,\bar{N}_{ abs}\,A_{l}}{g_l\, 8\pi^{3/2}\,\tilde{\nu}^3\,b},
\end{equation}
where $\tau$ is the optical depth of the \added{fundamental pure-rotational transition in the vibrational ground state, $\bar{N}_{abs}$ = $\log \bar{N}_{\rm abs}(^{12}CO)$, the column density seen in absorption \footnote{This is a fixed value not to be confused with parameter, $N_{\rm absorption}$, the number of rings fitted for absorption component for describing the absorption geometry.}}, $A_l$ is the Einstein A coefficient for the J=(1-0) transition, and $g_u$ ($g_l$) is the statistical weight of the upper (lower) state, and b is the intrinsic line width of the lines, taken to be \added{$0.15\,\mathrm{km\,s^{-1}}$}. 
We adopt a physically motivated, minimally broadened intrinsic line profile. The intrinsic line width is taken as the quadrature sum of thermal and turbulent components, \(b=\sqrt{v_{\rm th}^{2}+v_{\rm turb}^{2}}\). CO fluorescence modeling of Herbig Ae/Be disks indicates \(v_{\rm turb}\sim 2~\mathrm{km\,s^{-1}}\) \citep{Brittain+09}. 
% , and high-resolution spectroscopy at \(R\approx 50{,}000\) samples velocities at \(\sim 1.5~\mathrm{km\,s^{-1}}\) per pixel. 
For debris disks, where gas densities are low, we neglect turbulent broadening and adopt a thermally set \(b\). 
\added{For the cold $\sim\,45\,$K CO detected by ALMA, the thermal broadening is about $0.15 ~\mathrm{km\,s^{-1}}$. In addition to this physically motivated intrinsic width, we also account for the much larger instrument profile (e.g., Resolution $\sim 140~\mathrm{km\,s^{-1}}$).}
% Starting from a narrow intrinsic width preserves information and permits straightforward convolution to broader profiles (e.g., \(140~\mathrm{km\,s^{-1}}\)).
We only consider ground state transitions for the absorption comportment and take $u$ and $l$ to be 1 and 0. This absorption mainly suppresses the low-J v=1--0 emission.

Based on the UV flux of HD 131488 and oscillator strength, we can calculate the transition probability, allowing us to solve the fractional population of the ground and the first excited electronic state. The end goal of this calculation is to estimate the vibrational temperature of the gas at various distances in the disk to the star. These theoretically derived vibrational temperatures will be compared to the vibrational temperature of the gas fitted from phenomenological models to assess whether our models give a reasonable fit.  

We first calculate the band oscillator strength $f_{v'v''}$ following \citet{Beegle+99, Brittain+07} using equation 
\begin{equation}
    f_{v'v''} = \frac{(2-\delta_{0\Lambda^{'}})}{(2-\delta_{0\Lambda^{''}})} \frac{4\pi\epsilon_0 m_e c}{8\pi^2e^2}\,\lambda_{v'v''}^2 A_{v'v''}, 
\end{equation}\label{eq:oscillator}
where $\frac{(2-\delta_{0\Lambda^{'}})}{(2-\delta_{0\Lambda^{''}})}$ is the statistical weight of a transition and $\Lambda$ is the electronic angular momentum of the state. $v'$ and $v''$ denote the initial and final vibrational states and we only consider the fundamental transition where $\Delta v$ = |$v''$-$v'$| = 1. Similarly, as in equation \ref{eq:line-flux}, $A_{v'v''}$ is the Einstein A coefficient for transitions from the state $v'$ to the state $v''$. $\lambda_{v'v''}$ is the wavelength at which the transition happens, and the subscript follows the conventions of $A_{v'v''}$. $\epsilon_0$ is the absolute dielectric permittivity in vacuum, $m_e$ is the electron mass, $c$ is the speed of light, and $e$ is the electron \added{energy} in eV.
The transition probabilities are calculated following eq (4) in \citet{Brittain+07}
\begin{equation}
    g_{ij} = \frac{\pi e^2}{mc^2} \lambda^2 f_{ij} (\pi F_{\lambda})\, \textrm{photons}\,s^{-1} \textrm{molecule}^{-1},
\end{equation}\label{eq:transition_prob}
where $f_{ij}$ is calculated using eq \ref{eq:oscillator}. 
We write the theoretical fractional population of the ground electronic state following equation (5) in \citet{Brittain+07}
\begin{eqnarray}
     &\frac{dn_{X}^{v=i}}{dt} =  \sum\limits_{j} \Big( A_{A-X}^{v'=j, v''=i} + g_{A-X}^{v'=j, v''=i}\Big) n_{A}^{v=j} + A_{X-X}^{v'=i+1, v''=i} \nonumber\\
     & \times n_{X}^{v=i+1} - n_{X}^{v=i} \Big(\sum\limits_{j} g_{X-A}^{v''=i, v'=j} + A_{X-X}^{v''=j, v'=i} \Big) = 0 ,\label{eq:groundE}
\end{eqnarray}
which assumes the populations are in steady states and collisional excitations from other unseen molecules can be ignored and only $\Delta\nu=1$ transitions are important for ground electronic states. As shown in the schematic diagram in Fig. \ref{fig:schematic}, we ignore the IR fluorescence component because the transition probability from the X-A band ($g_{X-A}$) is much larger than the transition probability ($g_{X-X}$) due to IR fluorescence. 
We also write down the rate equation for the fractional population of the first excited states as 
\begin{equation}
    \frac{dn_{A}^{v=i}}{dt} = \sum\limits_{j} n_{X}^{v=i} g_{X-A}^{v'=i,v''=j} - n_A^{v=i} \sum\limits_{j}A_{A-X}^{v'=i,v''=j} = 0 \label{eq:1stE}
\end{equation}

\begin{figure}[h!]
    \centering
    \includegraphics[scale=0.41]{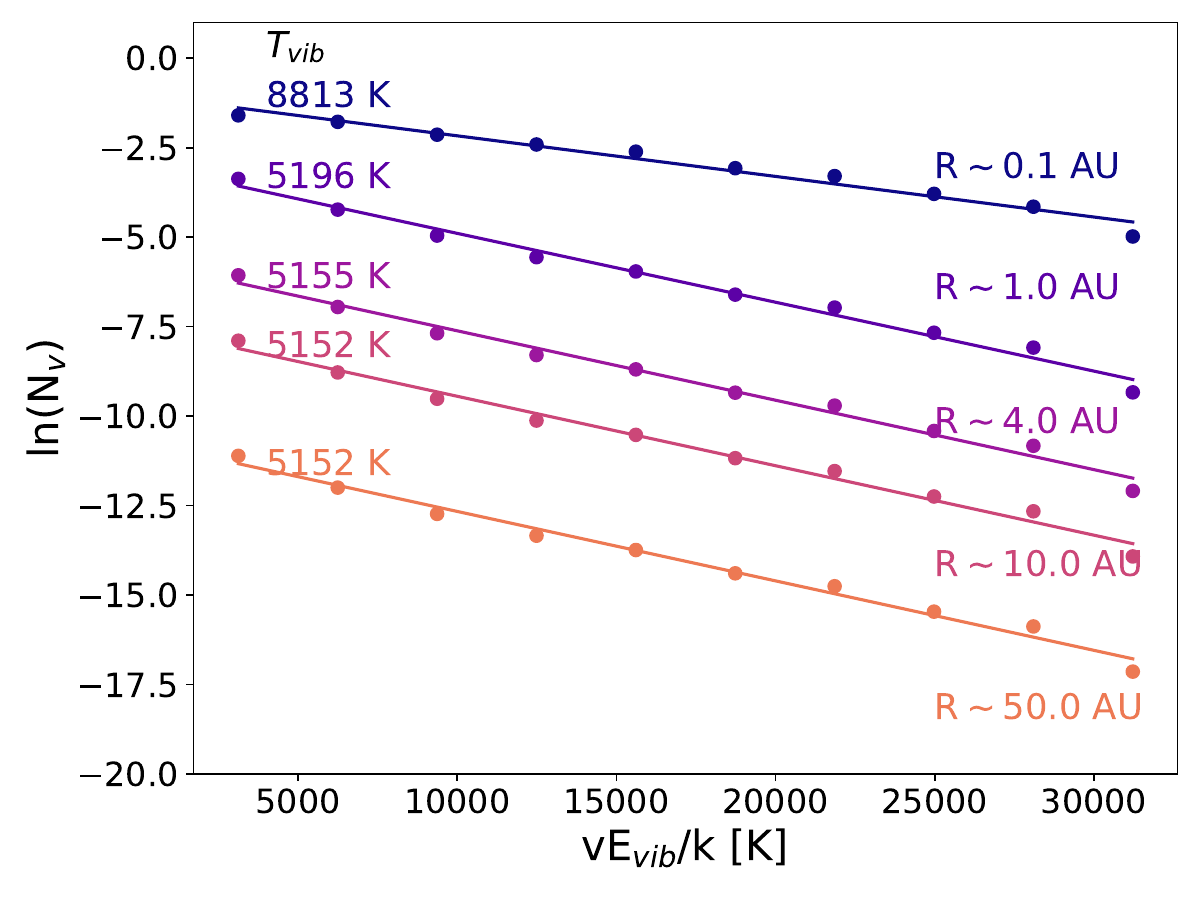}
    \caption{Fractional vibrational population excited by the radiation field of an A0 star at 0.1, 1, 4, 10, and 50 AU. \added{$E_{vib}/k$ for the x-axis represents the vibrational constant in units of Kevin. For the CO molecule, the vibrational constant is $3122$\,K. So for each vibrational level, $v$, the vibrational energy is $v\,\times\,3122\,$K.
    As the gas extends outward from \(0.1\,\mathrm{AU}\) to \(1\,\mathrm{AU}\), the stellar radiation field weakens and becomes more diluted. 
    Over this range, the fitted slope becomes progressively steeper, indicating that rovibrational transitions cool the vibrational levels more efficiently with increasing radial distances. 
    Beyond a few AU ($\sim4$\,AU), however, the slope changes only marginally, implying that the effectiveness of this vibrational ``cooling’’ saturates and remains nearly constant from \(4\,\mathrm{AU}\) out to \(50\,\mathrm{AU}\).}}
    \label{fig:vibPop}
\end{figure}

We calculate vibrational populations from v=0 to v=10 for the ground electronic state and v=0 to v = 9 for the first excited electronic state. Using equations \ref{eq:groundE} and 
\ref{eq:1stE}, we then have 21 coupled equations with 21 unknowns to solve for. We plot the fractional population level as a function of the vibrational energy for CO excited at 0.1, 1, and 10 AU in Fig \ref{fig:vibPop}.

\subsection{Considerations on Gas Geometry}\label{sub:geometry}

\begin{figure}[ht!]
\epsscale{2.6}
    \plottwo{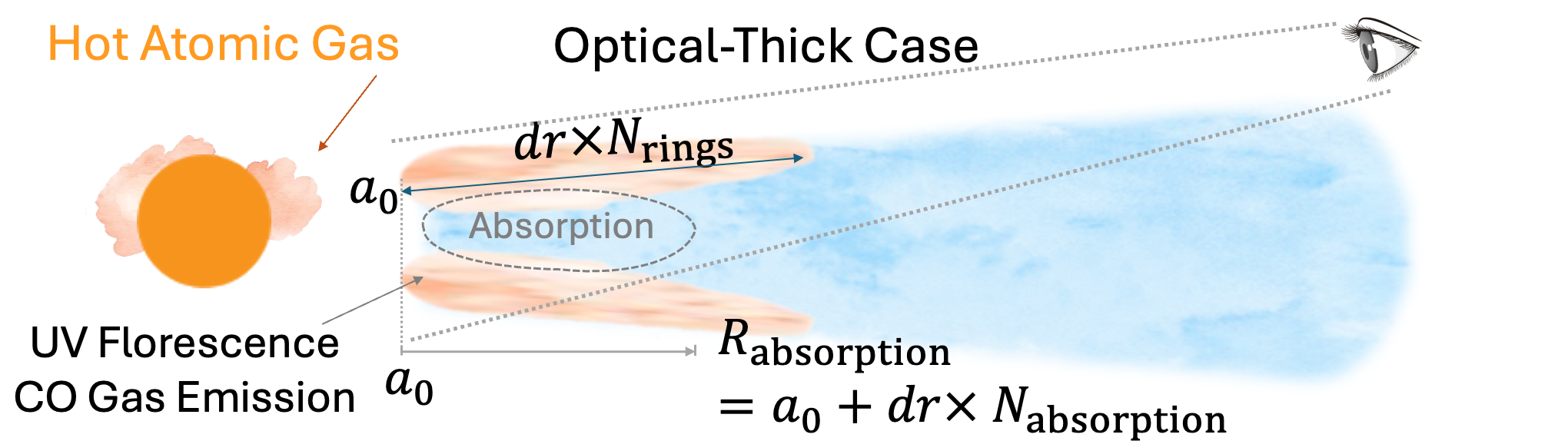}{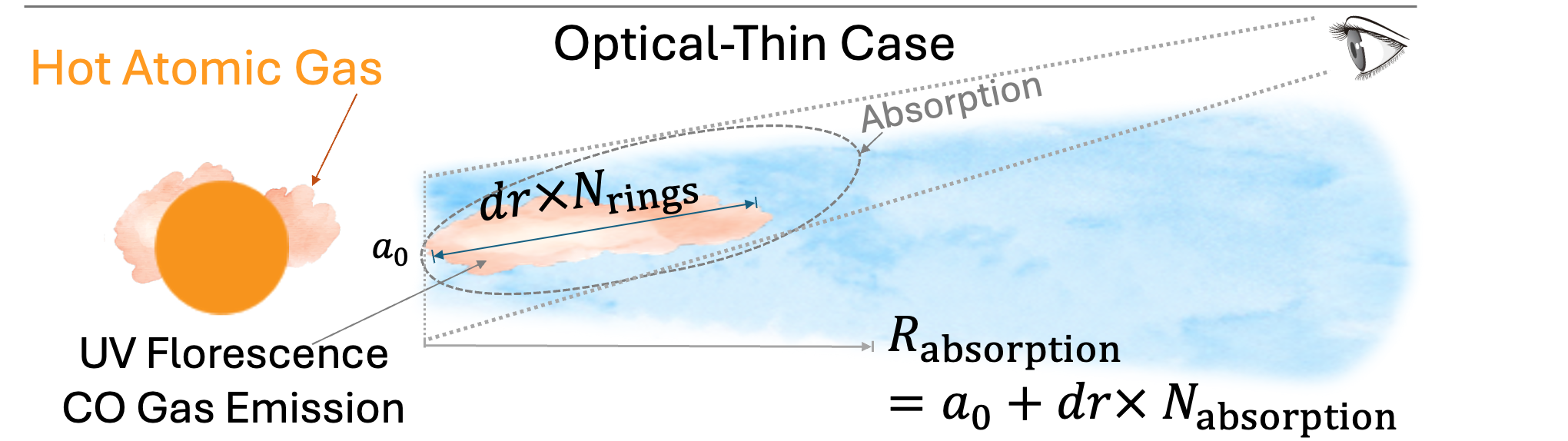}
    \caption{A Cartoon of the CO model for HD 131488. This diagram illustrates three key components, the hot atomic gas at $<0.1\,$ AU to the star \citep{Brennan+24, Rebollido+20}, the UV fluoresced CO gas modeled in annuli (this work), and the cold CO gas reservoir with a large spatial distribution (a few to $\sim100\,$AU) discovered with ALMA that creates the absorption signature seen in our JWST/NIRSpec data \citep{Moor+17}. For the newly discovered UV fluoresced CO gas component, $a_0$ is the inner gas radii, $dr$ is the width of each ring, and \added{$N_\text{rings}$} stands for the number of rings. The top and bottom panels show two scenarios, optical-thick and optical-thin, that are indistinguishable from JWST data due to a lack of spectral resolution to resolve the line profile. }
    \label{fig:cartoon}
\end{figure}

For simplicity, we assume that CO is distributed in ring-like annuli extending from an inner radius ($a_0$) to an outer radius ($a_\text{out}$), which is determined by the number of fix-width annuli, as illustrated in the cartoon by Figure \ref{fig:cartoon}. We fitted for the inner edge ($a_0$), and the number of rings ($N_{\rm rings}$), assuming the ring width (dr) is $0.1\,$AU wide from $0.1$ to $1\,$AU and increases to $1\,$AU wide outward of $1\, $AU. The ring width is $0.1$AU because this increment captures the radial gradient change in the gas vibrational temperature given the star's UV radiation. In the $1$ to $100\,$AU region, the gas vibrational temperature changes slowly with the radial distance, so $1\,$AU is sufficient to describe the change in the temperature changes as a function of radial distance. 

Within each annulus, the vibrational population of the CO molecules is set based on the CO gas vibrational temperature ($T_{vib}$), which depends on the radial distance ($r$) and stellar continuum emission around $1500$\AA. We visualize the vibrational population in Figure \ref{fig:vibPop}, where the vibrational populations for each $\Delta\,v$=$1$ transitions are plotted with solid circles. At the same radial distance, we fit the slopes of vibrational temperature based on the vibrational populations for all 10 levels of fundamental transitions. The inverse of the slope represents the $T_{vib}$, which we annotate on the plot near the lines. From top to bottom, we plot $0.1$, $1$, $4$ $10$, and $50$ AU in blue, purple, yellow, green, and red colors, respectively. \added{Figure \ref{fig:vibPop} x-axis can be understood as energy for each vibrational level. $E_{vib}/k$  is the vibrational constant in units of Kevin. For the CO molecule, the vibrational constant is $3122$\,K. So for each vibrational level, $v$, the vibrational energy is $v\,\times\,3122\,$K.
From $0.1$\,AU to $1$\,AU, the stellar radiation field weakens and becomes more diluted. 
Over this range, the vibrational temperature becomes increasingly lower (fitted slope becomes progressively steeper), which indicates that rovibrational transitions cool the vibrational levels more efficiently with increasing radial distances. This is because the relative populations of CO at v=1 versus v=2 decrease more rapidly for a steeper slope. 
Beyond a few AU, however, the slope changes only marginally, implying that the effectiveness of this vibrational ``cooling'' plateaus and remains nearly constant from $4\,$AU out to $50\,$AU.}

The second factor we added was power-law descriptions of the rotational temperatures. Following the results from HD~141569 and HD~100546 \citep{Brittain+03, Brittain+07}, we assumed that the gas extends from an inner radius ($a_0$) to an outer radius ($a_{out}$) and allowed the fiducial temperatures and power law indices to vary. We model the kinetic temperature of the CO gas with a broken-power law, primarily influenced by the radial distance from the central star and the gas temperature at the inner edge: 
\begin{equation}
    T_{\text{rot}} = T_{\text{rot}, 0} \left( \frac{r}{a_0}\right)^{\alpha} + T_{\text{base}}, 
\end{equation}
where $T_{\text{base}}$ represents the threshold temperature, which is set approximately equal to the kinetic temperature of the cold CO gas in absorption \citep{Brennan+24}. We require $T_{\rm rot}$ for gas in emission to be greater than $T_{\text{base}}$. Empirically, gas can remain warm through various mechanisms, such as collisional processes or viscous heating. To account for this uncertain heating mechanism, we set $T_{\text{base}}=45\,$K to ensure the newly detected gas population is not cooler than the known cold gas observed in absorption.

The final factor we considered is the effective radius of the cold CO foreground absorption. In the initial modeling, we assumed that the absorbing cold gas was perfectly co-located with the warm emitting gas. This assumption, however, led to an overprediction of the absorption line depth. Previous studies have shown that even a small misalignment ($\sim\,2\,\deg$) from an exactly edge-on disk orientation can result in non-detection of absorption in debris disks with ALMA CO detections \citep[e.g.,][]{Worthen+24}. Motivated by this, we refined our model by introducing an effective CO absorption radius, which significantly improved the fit. As illustrated in Figure~\ref{fig:cartoon}, the foreground gas component is offset from the warm gas geometry at a certain radial distance along the line of sight. We parameterize this offset with $R_{\textrm{absorption}}$, which is defined to be a region between a radius of $a_0$ to $R_{\textrm{absorption}}$ which gives rise to the absorption features to the model fit. The model flux for the absorption component is simply $F_{\rm abs} = \int_{a_0}^{R_{\rm absorption}}\pi r^2$ $F_{CO}(r) dr$. The spatial resolution elements for the absorption area are defined such that \added{$dr_{\textrm{absorption}}$} is taken to be $0.1$\,AU within $1$\,AU and $1$\,AU wide beyond $1$\,AU. In short, $N_{\textrm{absorption}}$ follows the same convention as $N_{\textrm{rings}}$.

The spectrum from each annulus was summed to produce the final spectrum. We fit for nine parameters, log(N($^{12}\text{CO}$)), Power law Index $\alpha$, $^{12}\text{CO}$/$^{13}\text{CO}$, $\text{T}_{\textrm{rot},0}$, $\text{T}_{abs}$, $\text{a}_{0}$, $\text{N}_{\text{rings}}$, $R_{\textrm{absorption}}$, and tabulate them in Table \ref{tbl:best-fit}. With these nine parameters, we build a model to fit the observed spectrum. Since we do not resolve the emission lines spectrally or spatially, there is a degeneracy between the surface area of the disk and the column density of gas we adopt. The optical depth of ro-vibrational CO emission resulting from UV fluorescence is generally optically thin and depends on the intensity of the radiation field, the flaring of the disk, and the gas/dust ratio of the disk \citep{Brittain+09}, so the mass of CO necessary to produce the observed emission is a lower limit on the total of CO in the disk. 

\subsection{Weighing Emission lines by their SNR and Model Optimization}\label{subsec:5sigma}
\begin{figure}[ht!]
\epsscale{1.25}
    \plotone{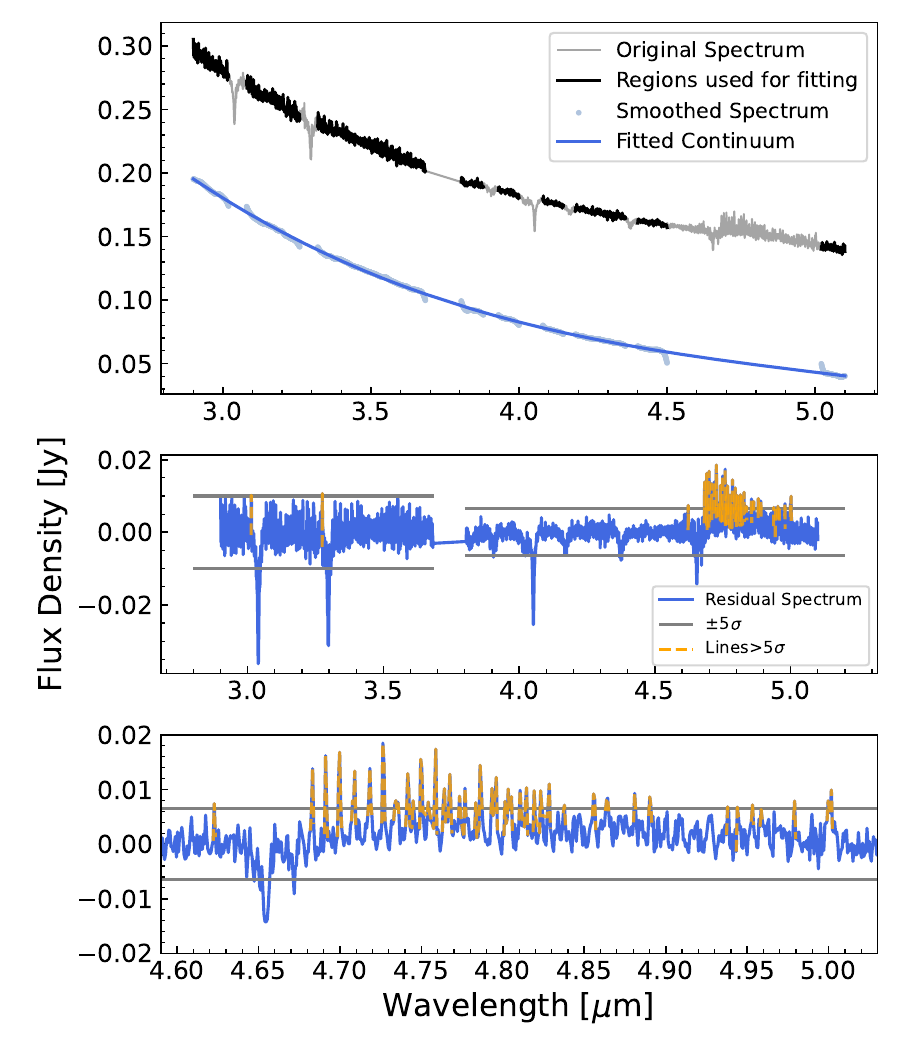}
    \caption{Identification of Lines above $5\sigma$. \textbf{Top}: The original JWST/NIRSpec spectrum (in gray) plotted against regions used for fitting (in black). The smoothed spectrum (light blue scatter) is offset from the original spectrum for illustration purposes. A fitted continuum (in solid blue lines is over-plotted on the smoothed spectrum. \textbf{Middle}: The residual spectrum after continuum subtraction. The $\pm\,5\,\sigma$ lines to the residual spectrum are plotted in gray. The lines with excess emission above $+\,5\,\sigma$ are overplotted in dashed orange lines and described in section \ref{subsec:5sigma}. Note that the spectral region with wavelengths shorter than $3.6\,\mu$m has larger spectral noises compared to that of the spectrum at wavelengths longer than $3.6\,\mu$m. \textbf{Bottom}:  A zoom-in of the $4.5$--$5.1\,\mu$m region to highlight CO (A-X) emission bands from the middle panel. }
    \label{fig:sigmaClip}
\end{figure}

We fit the spectrum by weighting each wavelength element by its local signal-to-noise ratio (SNR). Empirically, SNR weighting outperforms (i) weighting only selected emission lines and (ii) uniform weighting, both of which tend to overfit continuum-dominated regions while underfitting narrow emission features that occupy a small fraction of the wavelength range. In this subsection, we describe our approach in measuring the $5\sigma$ lines and assigning a weighing factor $\lambda$ to the $5\sigma$ lines relative to the rest of the spectrum. We also describe our exploration of the available model optimization functions to minimize the difference between the model and data.

We identify emission lines with an amplitude exceeding $5\sigma$ above the average continuum emission to minimize overfitting the regions dominated by noise. We first model the average continuum spectrum with a polynomial function. With visual inspections, we regard the following regions as noise-dominated ($2.9$, $3.02$), ($3.08$, $3.26$), ($3.26$, $3.32$), ($4.0$, $4.08$), ($4.15$, $4.19$), ($4.36$, $4.4$), and ($5.02$, $5.2$)$\,\mu$m. We visualize these regions in black solid lines in the top panel of Figure \ref{fig:sigmaClip}. We smooth the data in the noise-dominated (black line) regions with a Gaussian smoothing function implemented with \texttt{gaussian\_filter1d} from \texttt{scipy.ndimage} and $10\,\sigma$ for the gaussian kernel. We show the smoothed data in light blue dots offset from the original spectrum. We then use the B-spline function implemented with \texttt{BSpline} from \texttt{scipy.interpolate} with a smooth condition factor s=1, to fit the Gaussian-smoothed continuum regions. We show the resulting continuum in blue solid lines offset from the original spectrum in top panel of Figure \ref{fig:sigmaClip}. As the B-spline function passes through regions of interest with molecular emissions, we subtract the fitted continuum from the original spectrum to accentuate the emission features, which results in a residual spectrum. We show the residual spectrum in Fig \ref{fig:sigmaClip} middle panel with solid blue lines. We measure the standard deviations of the resulting residual spectrum  as our $1\,\sigma$ criterion, and plot the $5\,\sigma$ in solid, gray horizontal lines over the residual spectrum. 

We then identify the emission lines with an amplitude exceeding $5\sigma$ above the average continuum emission. Each identified emission line in our NIRSpec spectrum is at least defined by 3 points, which consist of one point for the line peak position, and the other two points defining the wings of the line. In some occasions of blended line profiles, there could be additional points for the wings of the lines. We identify the peaks that exceed the $5\sigma$ threshold and visualize the lines with these peaks in dashed orange color, as shown in the middle and bottom panel of Figure \ref{fig:sigmaClip}, resulting in 40 lines detected between $4.6$--$5\,\mu$m. 

We construct a custom function to optimize the model fit to the $5\sigma$ lines and minimize overfitting the rest of the spectra dominated by noise. We utilize the R2 score after empirically experimenting with commonly used metrics including \texttt{mean\_square\_log\_error, mean\_squared\_error, root\_mean\_squared\_error, max\_error, explained\_variance\_score, \textrm{and} r2\_score}. R2 score is advantageous at distinguishing models with severely overfitted noise features from models with slightly underfitted $5\,\sigma$ line regions. With the avoidance of overfitting in mind, we construct the minimization function (L) as below
\begin{equation}
    L = \frac{1}{R_2(>5\,\sigma)} - \lambda \frac{1}{R_2(<5\,\sigma)},
    \label{eq:minimization}
\end{equation}
where $\lambda$ is a weighting factor controlling the relative importance of regions below and above the $5\,\sigma$ threshold.
$\lambda$ is set to 5. 1/R$_2$ score is defined as
\begin{equation}
\frac{1}{R^2} = \frac{\sum_{i=1}^{N} (O_i - \bar{O})^2}{\sum_{i=1}^{N} (O_i - \bar{O})^2 - \sum_{i=1}^{N} (O_i - M_i)^2}, 
\end{equation}
where $O_i$ is the i$^{th}$ point in the observed spectrum and $\bar{O}$ is average value of the observed spectrum and $M_i$ is the corresponding model spectrum value. When computing $R_2$ score for lines above $5\,\sigma$ threshold, we set $O_i$ = $F^{obs}_{ >5\sigma,i}$, $M_i$ = $F^{model}_{>5\sigma,i}$, with a similar definition for regions below the threshold. Our optimization function L becomes negative when the model significantly overfits the noises in the spectrum and approaches positive values near unity as the fit improves. We present our best-fit results in section \ref{subsec:bestFitModels} using our custom optimization function with MCMC sampling. 

\subsection{Effect of Foreground ISM Absorption and Extinction on CO Features}
Since the foreground interstellar medium can create superimposed molecular absorption features on the spectrum, we examine whether the local ISM contributes to our observed CO features.
We check the column density along the line of sight towards HD 131488 to quantify whether local ISM contributes to our observed CO features. 
For line-of-sight towards HD 131488, the Galactic longitude and latitude are (l,b)  = $(326.67862^{o}, 15.978367^{o})$. According to \citet{Redfield_Linsky2000}, the velocity of the Colorado model of Local Interstellar cloud is $-20.3$ km $\cdot s^{-1}$ which corresponds to a wavelength shift of $\delta \lambda = 0.009~\mu$m at $4.5\mu$m where the CO fundamental emission features are located). Since the potential CO absorption signature from ISM will be significantly shifted out of the CO emission and absorption ($\delta \lambda = 0.009~\mu$m ) in the barycentric reference frame, the local ISM is unlikely to contribute to the observed features in the debris disk.
% , which have a factor of 6 narrower FWHM at $140$ km $\cdot s^{-1}$ ($0.00145~\mu$m). 
% a resolution of $\delta \lambda/\lambda = 0.0004$ (and
% have a much wider FWHM and therefore is 
\begin{figure*}[ht!]
\epsscale{1.1}
\plotone{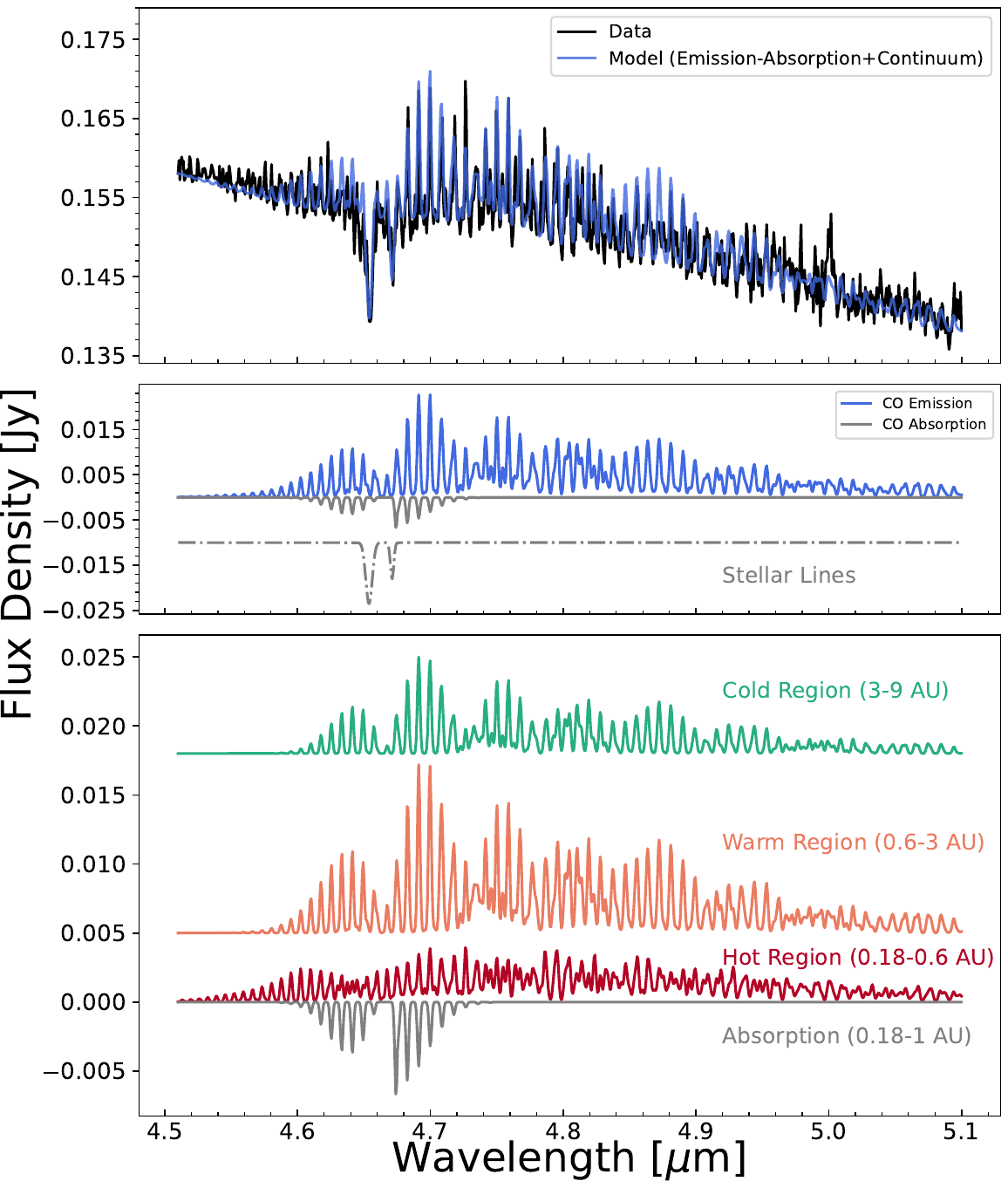}
\caption{\textbf{CO Model for HD 131488 with an Emission and an Absorption Gas Component.} \textbf{Top}: Best fit model (in blue) overplotted on NIRSpec data (in black). The best-fit model parameters are presented in Table \ref{tbl:best-fit}. \textbf{Bottom}: The decomposition of the model present in the top panel. The warm CO gas emission component is plotted in orange, and the cold CO in absorption is plotted in green. The stellar photosphere H lines (Humphrey and Pfund lines) are plotted in gray and offset for clarity. }
\label{fig:COmodel}
\end{figure*}
\section{Results}\label{sec4}
In this section, we first present the best-fit model, followed by a detailed discussion of the gas temperature profiles. We conclude by analyzing the properties of dust grains entrained in warm gas.

\subsection{Best-Fit Model}\label{subsec:bestFitModels}

We present our best-fit model in Figure \ref{fig:COmodel}, where the top panel shows the overall model alongside data, the middle panel shows the decomposition of the model into CO emission, CO absorption, and stellar absorption lines. The bottom panel shows a further decomposition of the CO emission model into three emission regions: the hot, warm, and cold regions.
\begin{deluxetable}{lcc}
\tablecaption{Best-Fit Model Parameters of the Median, 16$^{\rm th}$ and 84$^{\rm th}$ percentiles from MCMC posterior distribution.} \label{tbl:best-fit}
\tablehead{
\colhead{Parameter} & \colhead{Best Fit Value} & \colhead{Fit Range}}
\startdata
log(N($^{12}\text{CO}$)) & $14.45^{+0.23}_{-0.14} cm^{-2}$ & [12, 18]\\
 Power law Index $\alpha$& $-1.54^{+0.65}_{-0.75}$ & [-0.5, -4]\\
 for $T_{\text{rot}}$ & & \\
$^{12}\text{CO}$/$^{13}\text{CO}$ & $5.12^{+1.89}_{-1.93}$ & [1 ,81]\\
$\text{T}_{\textrm{rot},0}$ & $1155^{+84}_{-98}$ K & [80, 1400]\\
$\text{T}_{abs}$ & $39.12^{+4.22}_{-4.43}$ K & [30,60]\\
$\text{a}_{0}$ & $0.15^{+0.02}_{-0.02}$AU & [0.1, 10]\\
$\text{N}_{\text{rings}}$ & $18.07^{+2.17}_{-2.71}$ & [1, 100]\\
$N_{\textrm{absorption}}$  & $8.52^{+3.74}_{-5.00}$ & [1, $\text{N}_{\text{rings}}$] \\
$R_{\textrm{absorption}}$\tablenotemark{a}
  & $1.05^{+0.4}_{-0.5}\,$AU & [$\text{a}_{0}$, $\text{a}_{\text{out}}$] \\
% Residual & 0.0036\\
$\chi^2$ & 1150 & \nodata\\
$\chi^2_{\textrm{reduced}}$ & 1.01 & \nodata
\enddata
\tablenotetext{a}{Value calculated from $\text{a}_{0}$, $\text{N}_{\text{rings}}$ and $N_{\textrm{absorption}}$. Not a free parameter in the best-fit model. }
\end{deluxetable}\label{tbl:bestFit}

We set boundary conditions on parameters in our CO model (as described in section \ref{sub:CO-model}), leveraging knowledge from multi-wavelength observations (from FUV to mm) of HD 131488 and lending insights from gas emission properties in protoplanetary disks. Specifically, we have the following considerations:
\begin{enumerate}
    \item We require the cold CO population to have a consistent temperature as that measured from \textit{HST/STIS} UV absorption spectra \citep{Brennan+24}. Therefore, we allow $T_{abs}$ to vary between $30$ to $70$\,K and fix the column density along the line-of-sight to be N($^{12}CO_{abs})=10^{17}$, slighter lower than the HST value as the misalignment between disk PA and NIRSpec slit PA causes the full disk to extend outside of the slit. Fixing the N$^{12}(CO_{abs})$ value also helps with decreasing the degree of freedom in our model to speed up convergence. 
    \item We require the kinetic temperature ($T_{rot}$, approximated from gas rotational temperature) of the warm/hot CO population to range from $80$ to $1400$\,K, leveraging empirical knowledge of the CO emission temperature in the M band wavelengths from protoplanetary disks studies \citep{Brittain+07, najita2003, Salyk2011b, Vanderplas2015, vanderplas2009, Banzatti+2022}.  
    \item We let the ratio between $^{12}$CO and its isotopologue, $^{12}$CO/$^{13}$CO, range from $1$ to $81$, where $1$ represents an extremely isotope-rich environment and $81$ represents the higher limits to $^{12}$CO/$^{13}$CO ratio in the local ISM \citep{Milam+05,Ayres+13}. 
    \item For considerations on the geometry of the emitting CO population, we postulate that the warm gas must be located close to the central star and let the inner emitting radius $a_0$ to vary from $0.1$--$10$\,AU, let the total number of rings ($N_{rings}$) be a variable and fix the width of the rings to be $1$\,AU for rings exterior to $1$\,AU and $0.1$\,AU for rings interior to $1$\,AU to capture the change in the temperature radial gradient. The choice of ring width follows the common practice for modeling CO fluorescence gas in protoplanetary disks. We examine the 2D spectrum at the detector level and find that the CO lines are not spatially resolved compared to NIRSpec FS standard stars from the commission program (PID: 1128). Therefore, we limit the inner emission radius to be within $\leq\,15\,$AU, a resolution element for the beam of the JWST/NIRSpec FS observation. 
\end{enumerate} 

\begin{figure*}[th!]
    \centering
    \includegraphics[scale=0.7]{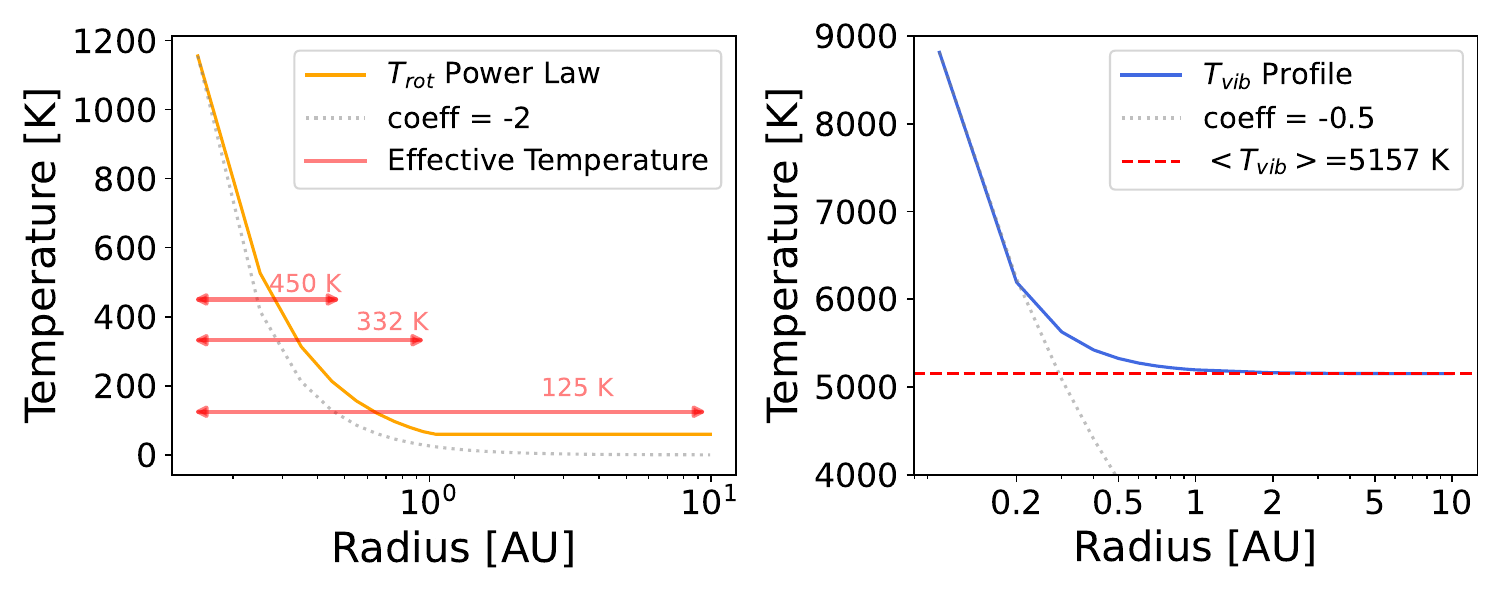}
    \caption{Gas Radial Distribution Profile. \textbf{Left}: Kinetic temperature of gas plotted as a function of radial distance (in orange solid lines, the power law index is $\alpha$=$-1.54$). The power law index of $-2$ is plotted in dotted gray line for comparison. The dashed horizontal red lines are the effective temperatures of gas averaged over $0.15$--$0.5\,$AU, $0.15$--$1\,$AU and $0.15$--$10\,$AU. \textbf{Right}: Vibrational temperature of the CO Gas. We plot the vibrational temperatures calculated from solving 21 coupled rate equations for an A0V type star in solid blue line. The power law index of $-0.5$ is plotted in dotted gray line for comparison. The vibrational temperature in the disk can be approximated by a shallow broken power with an index of $-0.5$. The dashed red horizontal line denotes the effective vibrational temperature across the entire emission region. }
    \label{fig:temp-profile}
\end{figure*}
With our initial conditions, we run our sampling algorithm and set a convergence criterion of $\chi^2_{\rm reduced}\leq1.5$, while using L (equation \ref{eq:minimization}) as our custom optimization function. We implemented the sampling algorithm with \texttt{emcee} \citep{emcee}. We use 30 walkers and $12,000$ iterations. We have a flat prior and we list the parameter prior ranges for each parameter in Table \ref{tbl:best-fit}. We also report our best-fit model parameters and their associated uncertainties in Table \ref{tbl:best-fit}. We show our MCMC corner plot in Figure \ref{fig:postDistribution} in Appendix A. We expand on the results in the following subsections.
\subsection{UV Fluorescence of Warm CO Gas} 
Analyses of the gas radial temperature profile indicate that the effective temperature for the emission region within $1$\,AU is close to terrestrial temperatures.
We take the inner edge gas rotational temperature ($\text{T}_{\textrm{rot},0}$) and power law index ($\alpha$) reported in Table \ref{tbl:best-fit}, and plot the gas radial profile. We show the results in Fig \ref{fig:temp-profile}, where the left and right panels present the kinetic and vibrational temperature profiles, respectively. In Fig \ref{fig:temp-profile} left panel, we calculate the effective temperature by weighting the temperature of each CO annulus emission and weighing them by their respective flux. We use this equation $\text{T}_{\text{eff}} = \sum_{i} F_{i}/F_{model} \times T_{i}$, 
where $T_{i}$ and $F_{i}$ are the respective kinetic temperature and flux of each CO annulus, and $F_{model}$ is the total flux. Interior to $0.5\,$AU, the $\text{T}_{\text{eff}}$ is $450\,$K, and interior to $1\,$AU, $\text{T}_{\text{eff}}$ is $332\,$K, indicating the bulk gas is terrestrial-like inwards of $1\,$AU. When we include the entire emission region interior to $10\,$AU, we obtain a $\text{T}_{\text{eff}}$ of $125\,$K, which is still much warmer than the cold ($60^{+9}_{-10}$K) CO gas \citep{Moor+17, Brennan+24}. 

Figure~\ref{fig:COmodel} illustrates the relative flux contributions from hot, warm, and cold regions. Spectral features between $4.5$ and $4.6\,\mu$m originate primarily from the hot component, while features from $4.6$ to $4.95\,\mu$m are composed of $\sim$25\% hot, $\sim$45\% warm, and $\sim$30\% cold gas. For highly excited vibrational lines between $4.95$ and $5.1\,\mu$m, all three regions contribute nearly equally. The detection of this temperature gradient in CO gas represents a new and significant result for debris disk studies.

The second key conclusion from our best-fit model is the confirmation of warm CO emission driven by fluorescent excitation. Our model shows a best-fit temperature of $T_{\rm rot}\sim\,1155\,$K and $T_{\rm vib}\sim\,8800\,$K at the inner edge of warm CO emission.  
The high vibrational temperature $T_{\rm vib}$ of gas is close to the diluted color-temperature of the stellar radiation field and does not represent the kinetic temperature of the newly discovered warm gas. The rotational temperature of the warm gas is a proxy for the true thermal temperature of the gas, and the gas flux-weighted average temperature is close to terrestrial temperatures. As shown in Fig \ref{fig:temp-profile}. This large difference between $T_{\rm vib}$ and $T_{\rm rot}$ indicates that the emitting CO population is not in local thermal equilibrium (LTE). CO molecules first absorb energy from incident UV photons and are excited to higher electronic states. Then, vibrational relaxation makes the excited state electrons go to a lower energy level and eventually to the ground state.

Signatures of UV fluorescence in warm gas are evident from individual spectral lines that are both vibrationally and rotationally highly excited. As shown in Figure~\ref{fig:examples} (ROI~3), numerous lines in the $4.8$--$5.1\,\mu$m region exhibit high $v$ and high $J$ transitions, including v(7–6) J(17–16), v(4–3) J(6–5), and v(6–5) J(7–6) of $^{12}$CO. Additional $^{13}$CO ro-vibrational lines with high $v$ and $J$ are also present (blue shading) but not individually annotated to avoid clutter. The data underlying Figure~\ref{fig:vib-Levels} are provided for reference. These detections confirm UV fluorescence as the excitation mechanism for warm CO gas in HD~131488.

\subsection{Properties of Gas and Dust in the Terrestrial-Planet Forming Zone}\label{sub:gasimplications}
In this subsection, we estimate the mass of micron-sized dust grains entrained in newly discovered warm CO gas based on the measured $^{12}$CO/$^{13}$CO ratio. JWST is uniquely sensitive to this population of close-in dust grains, which ground-based coronagraphic imaging is insensitive to. Our calculations yield a dust mass of $\sim 5 \times 10^{21}$\,kg, approximately $20$ times the mass of the asteroid Vesta. This warm dust population is also about 20 times more massive than the dust associated with the recent giant collision identified in the terrestrial zone of the $\beta$~Pic planetary system \citep{Chen+2024}.

The $^{12}$CO/$^{13}$CO ratio in transition disks is used as a measurement of dust opacity entrained in the gas. \citet{Brittain+09} suggest that the $^{12}$CO/$^{13}$CO ratio should be 1 when UV fluoresced gas is free of dust. This is because $^{12}$CO becomes optically thick to UV much sooner than $^{13}$CO based on rate coefficients and oscillator strength. Therefore, $^{13}$CO can continually be excited to a larger column density while the excited $^{12}$CO column density remains constant. Eventually, the $^{12}$CO/$^{13}$CO ratio becomes unity because we are probing different column densities in $^{12}$CO and $^{13}$CO respectively. A $^{12}$CO/$^{13}$CO ratio of larger than 1 implies the existence of dust entrained in the upper atmosphere of the gas. Dust is a continuum opacity source in near UV wavelengths at 
shorter than 1500$\,\text{\r{A}}$ ($0.15\,\mu$m hereafter) that limits the maximum column density of $^{12}$CO and $^{13}$CO excited by the UV photons from the stellar radiation field. As reported in Table \ref{tbl:bestFit}, our best fit best-fit values of $^{12}$CO/$^{13}$CO ratio is around $5$ with an uncertainty of $1.9$, which we can use to estimate a dust column density entrained in the warm gas.

We use the best-fit values of $^{12}$CO/$^{13}$CO ratio to constrain the column density of dust grains that are optically thin to UV photons and entrained in the surface layers of the newly discovered warm gas. 
The dust opacity is determined by its optical constants in the UV wavelength range ($1100$--$1500\,\text{\r{A}}$). We take the extinction coefficients ($Q_{\rm ext}$) around $1100$--$1500\,\text{\r{A}}$ to calculate $\tau_{\rm dust}$. Using $\tau_{\rm dust} = e^{-Q_{\rm ext}}$, we obtain the value to be around $1$ to $2$, depending on the dust grain composition. The $\tau_{\rm dust}$ has a very weak wavelength dependence for these UV wavelengths. Given that HD 131488 is known to have carbonaceous grains \citep{Melis+13}, we use PAH-carbonaceous grains data from \citep{Li+Draine2001} \footnote{Opacity data used is  \href{https://www.astro.princeton.edu/~draine/dust/dust.diel.html}{here}} and round $\tau_{\rm dust}$ to $1$ for simplicity.
We first estimate the column density of dust needed to reach an opacity of unity ($\tau$=$1$). We follow the methods in \citet{Laor+93} using the equation

\begin{equation}
\tau_{\text{dust}} = \int^{a_{max}}_{a_{min}} Q_{ext}\pi a^2 n_0 \frac{dn}{da} da,  
\label{eq:tau}
\end{equation}
where $a_{max}$ (and $a_{min}$) are the maximum and minimum grain sizes taken to be $0.1$ and $10\,\mu$m. $Q_{ext}$ is extinction coefficient calculated at $0.15\,\mu$m, $n_0$ is scale factor for the dust number column density, $\frac{dn}{da}$ is the grain size distribution following \citet{Dohnanyi+69} of $a^{-3.5}$. We solve for $n_0$ using equation \ref{eq:tau} by setting the left-hand side of the equation to 1. We obtain $n_0$ = $1.75 \times 10^{-2} \mu\textrm{m}^{1/2}$ using optical constants of ISM-like astrosilicates \citep{Laor+93}. 
We then substitute $n_0$ into the following equations to calculate the mass column density. 
% We calculate the dust number column density $n_0$ using 
% \begin{equation}
%     N = \int^{a_{max}}_{a_{min}} n_0 \frac{dn}{da} da, 
% \end{equation}
and the dust mass column density M
\begin{equation}
    M = \int^{a_{max}}_{a_{min}} n_0 \frac{dn}{da} \rho \pi a^3 da, 
\end{equation}
where $\rho$ is the density of silicate dust grains and is taken to be $3.3\,g\cdot cm^{-3}$. We get a mass column density of $8.8 \times 10^{-5}\,g \cdot cm^{-2}$, corresponding to number column density of $2.2 \times 10^{8}\,cm^{-2}$, to reach an optical depth of 1 at $0.15\,\mu$m. We tabulate these resulting values in Table \ref{tbl:dust-prop}. 

\begin{deluxetable}{lcc}
\tablecaption{Properties of Dust}
\tablehead{
\colhead{Parameter} & \colhead{Value}}
\startdata
Dust number column density N &  $2.2 \times 10^{8}\,cm^{-2}$ \\
Dust mass column density M &   $8.8 \times 10^{-5}\,g \cdot cm^{-2}$ \\
Dust mass Estimate &  $5 \times 10^{24}\,g$ ($\sim\,20\,\text{M}_{\rm Vesta}$) \\
N(warm Dust)/$N(^{12}$CO, warm) & $2\times 10^{-6.45}$\\
N(warm Dust)/$N(^{13}$CO, warm) & $1\times 10^{-5.45}$\\
M(warm Dust)/M(CO, warm)& $6.7 \times 10^{3}$\\
\enddata 
\label{tbl:dust-prop}
\end{deluxetable}

We then constrain the location of gas and the dust entrained in the gas in the disk and find that we are sensitive to dust at least warmer than $\sim\,332\,$K, similar to terrestrial temperature. To estimate the equivalent location of dust, we adopt the average temperature of $450$, $332$, and $125\,$K estimated from our best-fit gas kinetic temperature profile for the dust grains assuming that these dust grains are in radiative equilibrium where the energy received by a dust grain equals to the energy it re-emitted, such that 
\begin{eqnarray}
    & E_{in} = E_{out}\nonumber\\
    & P_{in} \cdot \pi\,r_{g}^2 = P_{out} \cdot 4\pi\,r_{g}^2 \nonumber \\
    & \frac{1}{4} P_{in} \cdot A_{grain} = P_{out} \cdot A_{grain} \nonumber\\
    & \pi r_{g}^2 \int_{0}^{\infty} j_{\lambda}(T_{\star}) Q_{abs} d\lambda = 4\pi r_{g}^2 \int_{0}^{\infty} Q_{em} B_{\lambda}(T_{grain}) d\lambda, \nonumber\\
        & \int_{0}^{\infty} j_{\lambda}(T_{\star}) Q_{abs} d\lambda = 4 \int_{0}^{\infty} Q_{em} B_{\lambda}(T_{grain}) d\lambda,
\end{eqnarray}
where $j_{\lambda}(T_{\star})$ is the specific intensity of the star, $Q_{abs}$  and $Q_{em}$ are the absorption coefficient and emissivity of the dust grain, and $B_{\lambda}(T_{grain})$ is spectral radiance of the dust grain. $P_{in}$ and $P_{out}$ are the power received and re-radiated by the grain. $A_{grain}$ stands for the surface area of the grain. $T_{\star}$ and $T_{grain}$ are the temperatures of the star and the dust grain. 
We can then solve for the grain distance based on their temperature such that 
\begin{equation}
    d_{\textrm{grain}} = \frac{R_{\star}}{2}\Big(\frac{ T_{\star}}{T_{\text{grain}}} \Big)^{(2+0.5\,p)}.
\end{equation}
In the case of grain acting as a blackbody, $Q_{abs}$ and $Q_{em}$ are unity as blackbody radiation is independent of the incident wavelength and p\,=\,$0$.  For an A0V type star with a  $T_{\star}=8800\,K$ and $R_{\star}=2.2\,R_{\odot}$, we obtain a distance $d_{\textrm{grain}}$ of $\sim 2.0$, $3.6$ and $15.5\,$AU for large grains ($a \geq  100\,\mu$m) or a lower limit to the distance at which small sub-micron-sized dust grains are located compared to the large grains. In the small grains limit where $2\pi\,a/\lambda>>1$, their emissivity follows $Q_{abs} \propto (1/\lambda)$ and $p\,=\,1$, according to \citet{Jura+93}. Therefore, small, sub-micron-sized dust grains are approximately located at $\sim 8.7$, $18.5$, and $114.8$ \,AU from the star. Since the gas we observe is not spatially-resolved in the beam of NIRSpec FS observation, we believe the gas is most likely from regions inward of $15\,$AU from the host star. Therefore, it is unlikely we are sensitive to grains out to $114\,$AU. The grains are mostly likely warmer than the gas temperature, and at least warmer than $\sim 332$K. Comparing our estimated dust mass and location with ground-based coronagraphic studies, we find that JWST probes dust within the region occulted by the coronagraph, independent of assumptions on dust properties.
\begin{deluxetable}{lcc}[h]
\tablecaption{Estimated Locations of Dust Entrained in Warm Gas}
\label{tbl:dust-prop}
\tablehead{
\colhead{T$_{\text{grain}}$ [K]} & \colhead{$d_{\text{grain, lg}}$ [AU]}& \colhead{$d_{\text{grain, sm}}$[AU]} \\
& \colhead{Large grains}& \colhead{Small grain limit}\\
& \colhead{($a \geq  100\,\mu$m)}& \colhead{ (2$\pi$a$<<\lambda$)}}
\startdata
$450$ & 2.0 & 8.7\\
$332$ & 3.6 & 18.5\\
$273$\tablenotemark{*} & 5.3 & 30.2 \\
$125$ & 15.5& 114.8 \tablenotemark{a}\\
\enddata
\tablenotetext{a}{Since the CO gas emission lines in our NIRSpec Fixed Slit (FS) observations are not spatially resolved within the telescope beam, our data primarily probe regions within  $\sim\,15\,$AU of the central star. This spatial resolution limit makes it unlikely to detect dust grains at distances as large as $114\,$AU.}
\tablenotetext{*}{We calculate the equivalent distance of $273\,$K grain in thermal equilibrium to approximate the location of the water-ice line.}
\end{deluxetable}

We further approximate the location of the water-ice line to get an idea of whether these grains can be icy or desiccated. 
We consider warm dust grains in both small and large grain limits
in comparison with the equilibrium distances of a $273$\,K grain ($\sim 5.3$\,AU for large grains and $\sim 30.2$\,AU for small grains) 
Since the entrained warm grains lie at temperatures above the water-ice freeze-out, they are most likely located inside the water-ice line, a region inaccessible to ground-based coronagraphy. Comparing these distances with the peak locations of the warm and cold belts at $\sim 10$ and $\sim 100$\,AU from SED, respectively, in Fig.~15 of \citet{Pawellek+24}, we conclude that the warm dust and gas are consistent with an origin within the warm belt. Furthermore, prior SED analyses from previous studies \citep{Melis+13, Lisse+17} revealed a hot dust excess at $3$--$4\,\mu$m, corresponding to populations at $\sim 6$ and $\sim 35$\,AU. The warm CO gas may therefore be co-located with the $\sim 6$\,AU dust component.

\begin{deluxetable}{lcc}[h]
\tablecaption{Properties of Gas}
\label{tbl:gas-prop}
\tablehead{
\colhead{Parameter} & \colhead{Value}& \colhead{Reference}}
\startdata
$\bar{\text{T}}_{\text{rot}}$  $\leq\,0.5\,$AU & $450$ K & This Work\\
$\bar{\text{T}}_{\text{rot}}$ $\leq\,1\,$AU & $332$ K & This Work\\
$\bar{\text{T}}_{\text{rot}}$ $\leq\,10\,$AU & $125$ K & This Work\\
Average $\text{T}_{vib}$ & $5157$ K& This Work\\
$\text{M}(\text{warm}^{12}\text{CO})$ & $7.5^{+5.2}_{-2.1}\times 10^{20}$ g& This Work\\
& (or $1.25^{+0.87}_{-0.34} \times10^{-7}M_{\oplus}$)& \\
$\text{M}(\text{ALMA}^{12}\text{CO})$ & $(8.9\pm 1.5)\times10^{-2}M_{\oplus}$& \citet{Moor+17}\\
\enddata 
\end{deluxetable}

% \begin{deluxetable}{lcc}[h]
% \tablecaption{Estimated Locations of Dust Entrained in Warm Gas}
% \label{tbl:dust-prop}
% \tablehead{
% \colhead{T$_{\text{grain}}$ [K]} & \colhead{$d_{\text{grain, lg}}$ [AU]}& \colhead{$d_{\text{grain, sm}}$[AU]} \\
% & \colhead{Large grains}& \colhead{Small grain limit}\\
% & \colhead{($a \geq  100\,\mu$m)}& \colhead{ (2$\pi$a$<<\lambda$)}}
% \startdata
% $450$ & 2.0 & 8.7\\
% $332$ & 3.6 & 18.5\\
% $273$\tablenotemark{*} & 5.3 & 30.2 \\
% $125$ & 15.5& 114.8 \tablenotemark{a}\\
% \enddata
% \tablenotetext{a}{Since the CO gas emission lines in our NIRSpec Fixed Slit (FS) observations are not spatially resolved within the telescope beam, our data primarily probe regions within  $\sim\,15\,$AU of the central star. This spatial resolution limit makes it unlikely to detect dust grains at distances as large as $114\,$AU.}
% \tablenotetext{*}{We calculate the equivalent distance of $273\,$K grain in thermal equilibrium to approximate the location of the water-ice line.}
% \end{deluxetable}

\subsection{CO Mass Estimate and Collisional Partners}\label{sec:collision-partners}
\begin{figure*}[ht!]
\epsscale{1.4}
    \plotone{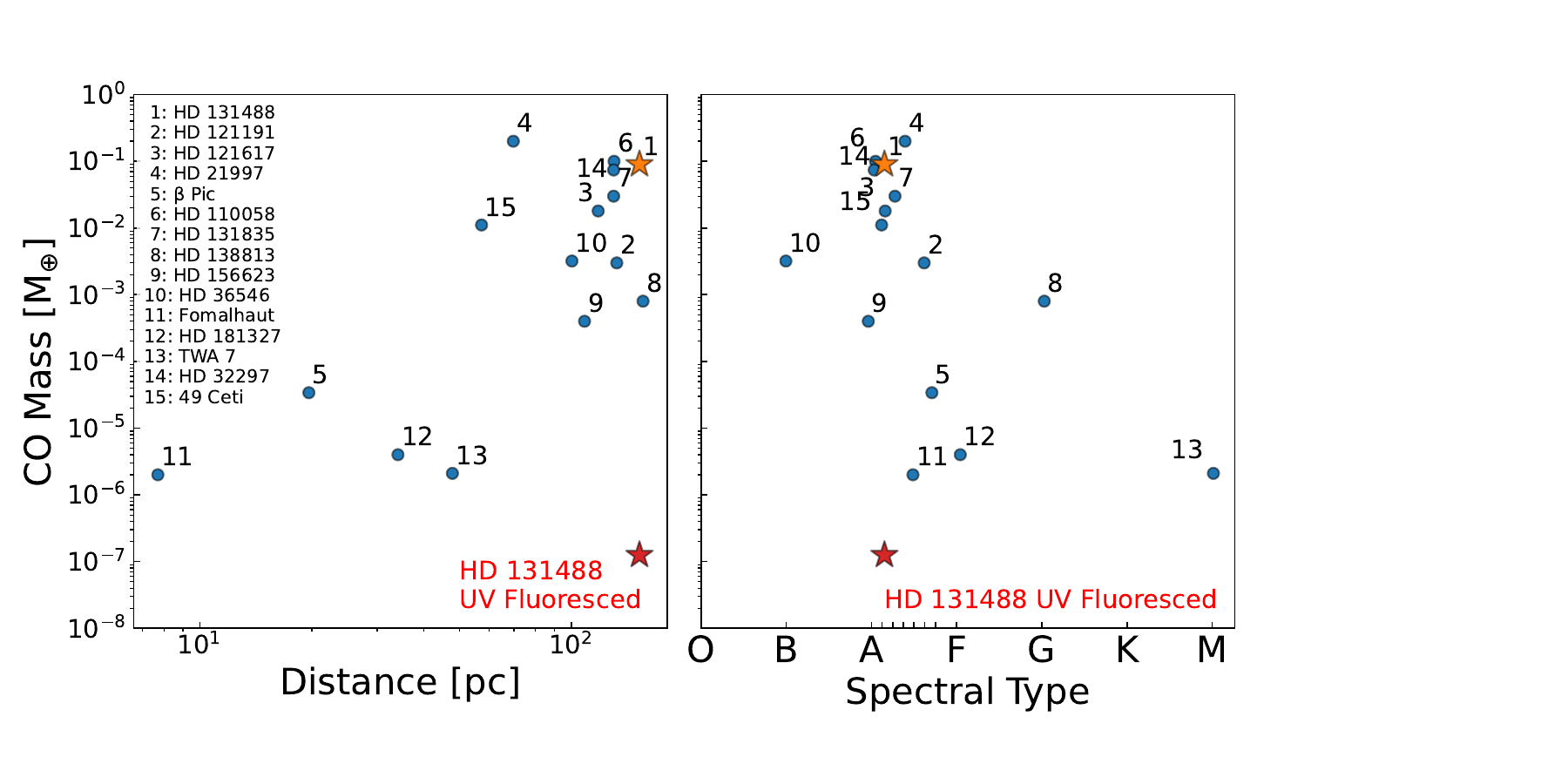}
    \caption{Comparison between HD 131488 Fluoresced Gas and ALMA detected CO gas in Debris Disks. \textbf{Left}: CO mass in debris disks as a function of distance to Earth with our estimated CO masses of the UV fluoresced gas components for HD 131488 (red star). HD 131488 is denoted as a star, and other disks are denoted with a circle. \textbf{Right}: CO mass in debris disks as a function of stellar types for the host stars. The warm CO gas is denoted as a star. }
\label{fig:comparison}
\end{figure*}

An estimation of CO mass reveals that the NIRSpec spectrum is sensitive to CO at one part in 10 millionths of an Earth mass. We estimate the CO mass using the equation: 
\begin{equation}
    \text{M}_{\text{CO}} = \text{N}(^{12}CO)\times \text{A}/A_V \times m(CO), 
\end{equation}
where $\text{A}$ is the emitting area defined as $\pi \times (R_{out}^2-R_{in}^2)$, $A_V$ is the Avogadro's number, and m(CO) is the molar mass of CO taken to be $28.01$ g/mol. We obtain $7.5^{+5.2}_{-2.1}\times 10^{20}$ g (tabulated in Table \ref{tbl:gas-prop}), which is approximately one part in 10 millionths of the Earth's mass. 

Compared to ALMA-discovered CO masses, UV fluorescence probes CO masses at a significantly lower column density, a factor of a million lower than previous CO detection for the same disk, and a factor of 10 more sensitive to the lowest CO mass discovered in debris disks to date. As plotted in Figure \ref{fig:comparison}, we show the UV fluoresced gas of HD 131488 in red star in comparison with the CO masses of debris disks derived from ALMA studies \citep{Marino2016, Moor+17, Moor2019, Cataldi2023, Higuchi2020, Rebollido+22, Matra2017, Matra2019, Hales2022}. In Fig. \ref{fig:comparison} left panel, we plot the CO mass as a function of distances to Earth \citep[Gaia DR2; ][]{Bailer-Jones2018}. HD 131488 is denoted with a star symbol, where the warm UV fluoresced component is in red and the cold CO is denoted in orange. Fig. \ref{fig:comparison} right panel shows CO mass as a function of stellar types sourced from the Simbad databases \citep{Wenger2000}.  Both panels reveal that UV-fluoresced CO probes lower column densities than ALMA detections, yielding smaller masses for tenuous debris disk gas. In HD 131488 specifically, the UV-fluoresced component has $\sim 10^5\times$ lower column density than the cold CO measured by ALMA. Compared to all known CO-bearing debris disks, the UV-fluoresced mass in HD 131488 is still $10\times$ smaller than the previous minimum (Fomalhaut).

CO molecules can be excited from various mechanisms, including UV continuum photons ($8$--$13$eV) directly pumping electronic states, IR continuum ($1$--$3\,\mu$m) photons ($0.4$--$1.2$eV) exciting vibrational overtones, collisional energy transfers from other molecules, and fluorescence following UV photodissociation of volatiles (i.e. H$_2$O, CH$_4$), depending on many factors including gas density, dusty extinction, strength of the ambient UV field, and CO abundance \citep{Brittain+07, najita2003, Salyk2011b, Vanderplas2015, vanderplas2009, Banzatti+2022, Brittain2023, boney2013infrared}. Our discussion focuses exclusively on collisional excitation by unseen partners (e.g., H$_2$ and H$_2$O). We acknowledge that other processes may contribute significantly to the excitation budget.

\subsubsection{H$_2$ as a Collisional Partner}
We estimate the upper limits for collisional partner column densities required to pump CO to excited states. We show that with the assumptions detailed below and without accounting for the dissociation of H$_2$ into atomic hydrogen, collisional excitation for CO by unseen H$_2$ remains a plausible mechanism to account for the warm CO emission.
We begin with a molecule commonly observed in protoplanetary disks but yet to be observed for debris disks, H$_2$. We investigate the possibility of H$_2$ acting as a collisional partner to the newly discovered warm CO gas. 
Assuming thermal equilibrium, $T_{\text{rot}}$=$T_{\text{kin}}$, the critical number density for rotational thermalization is
\begin{equation}
    n_{\text{crit}} = \frac{A_{ul}}{C_{ul}}, 
    \label{eq:collisional-partners}
\end{equation}
for fundamental transitions from J=$i\rightarrow\,(i-1)$, with i from 1 to 40. 
$A_{ul}$ is the Einstein coefficient from the upper state $u$ to the lower state $l$ and in units of $s^{-1}$. Similarly, $C_{ul}$ is the collisional rate between CO and another molecule (in our case, CO-H$_2$ or CO-H$_2$O determined experimentally) from the upper state $u$ to the lower state $l$ in units of $cm^{3}\cdot\,s^{-1}$.

From our best-fit model, we find that J=$\,13\rightarrow\,12$ is the highest J transition detected for ground state vibrational level v(1-0), and we denote the line in Figure \ref{fig:highJ} in the appendix. 
We assume J=$\,13\rightarrow\,12$ transition is thermalized due to collisions with another unseen molecule, and calculate the critical number density of that unseen molecule. 
% We find the critical number density at J=$\,13\rightarrow\,12$ where $N(CO)$<$n_{\text{crit}}$. 
Therefore, if we assume the H$_2$/CO ratio of $10^4$, we infer that CO rotational lines could be thermalized to at least J =$13$ for a gas population with an average rotational temperature of $300\,$K.
% The critical number density to thermalize these rotational levels can be estimated from the ratio of the Einstein A coefficient and collisional rate.  
We caution readers that rigorous analysis requires spectrally resolved CO lines, but NIRSpec's blended lines necessitate this assumption for estimation.
Assuming the collisional partner is molecular hydrogen H$_{2}$, the upper limit to H$_{2}$ volume density can be estimated by 
\begin{equation}
    n(\textrm{H}_2)\times C_{\textrm{CO-H$_2$}} = A_{13, 12}.
\end{equation}
Rates for the collisional excitation of CO by $H_2$ are given by \citep{Yang2010}. For the J=13 transition, the Einstein A coefficient $A_{13,12}$ is  $1.7 \times10^{-4} s^{-1}$ and the collisional rate at $330\,$K is $3.8 \times10^{-11} cm^{3}s^{-1}$ , thus the critical volume density of the H$_{2}$ gas is $\sim\, 4.5\times10^{6} \textrm{cm}^{-3}$.

To estimate the column density of  H$_{2}$ gas, we boldly assume H$_{2}$ has a similar spatial distribution to the warm CO gas and show details of the calculation for a lower limit to H$_{2}$ gas mass in Appendix C.
We obtain a 2D column density upper limit of $1.7 \times10^{19} cm^{-2}$ to $4.5 \times10^{21} cm^{-2}$ depending on assumptions. Therefore, the lower limit to the column density ratio of CO/H$_{2}$ is $5 \times 10^{-8}$ to $10^{-5}$, and a mass ratio of $7 \times 10^{-7}$ to $10^{-4}$. Take the mass ratio on the lower end, we obtain an H$_{2}$ gas mass upper limit of $7.5\times10^{24}$g, equivalent to $1\times10^{-3}~\text{M}_{\oplus}$.  Future JWST/MIRI observations targeting H$_2$ could further test whether CO–H$_2$ collisions alone can account for the observed warm CO emission.

\subsubsection{H$_2$O as a Collisional Partner}
Motivated by the possibility that the newly observed warm CO originates from exocomets, we investigate whether outgassed H$_2$O in cometary comae can serve as a collisional partner for the CO gas. If the warm CO gas arises entirely from collisional release of CO frost in planetesimals (or known as ``secondary origin''), H$_2$O emerges as a plausible collisional partner. In solar system cometary comae, collisional excitation is dominated by H$_2$O at smaller heliocentric distances and by CO at large heliocentric distances. Furthermore, local thermal equilibrium (LTE) for the collision between CO and H$_2$O particles can be achieved in high-density, or dust-rich cometary comae at small heliocentric distances (a few AU for the solar system) \citep{Espinasse+1993, Squyres+1985}. Therefore, for simplicity, we assume local thermal equilibrium (LTE) for the collision between CO and H$_2$O particles in our calculations, because debris disks are known to be dust-dominated planetary systems, and our detected warm CO is at a close-in (a few AU) distance to the star. 

We estimate the upper limits to column density of H$_2$O following equation \ref{eq:collisional-partners}. For convenience, we assume that, similar to using H$_2$ as a collisional partner, the critical number density for H$_2$O might be thermalized at J=$\,13\rightarrow\,12$ where $N(CO)$<$n_{\text{crit}}$.
The Einstein A coefficient (A$_{13, 12}$)is an intrinsic property of the CO molecule and independent of gas temperature, and is therefore $2.5 \times10^{-6} s^{-1}$, same as used in calculation with H$_2$. The CO--H$_2$O collisional rate is a function of gas temperature. The CO--H$_2$O collisional rate has been measured in solar systems with the temperature range of $10$--$100$K. We adopt the rates K$_{\rm CO-H_2O}=8\times10^{-11}\textrm{cm}^{3}\cdot s^{-1}$ at $100$K of the thermalize population of CO and H$_2$O molecules from \citet{Faure+20} \footnote{Note that para- and ortho-H$_2$O has similar collisional rates.}. The quoted kinetic temperature for this collision rate is roughly consistent with the bulk temperature of warm gas averaged over the inner $10\,$AU from our model. We obtain an upper limit to the critical density of H$_2$O of $3.1\times10^{4} \textrm{cm}^{-3}$, discounting other possible excitation mechanisms. To estimate the upper limit mass of H$_{2}$ gas, we assume that the H$_{2}$ has a similar spatial distribution as the warm CO gas. We similarly estimate an upper limit for water vapor mass of $m(\mathrm{H_2O}) = 9.9 \times 10^{22}$ g ($1.6 \times 10^{-5} M_\oplus$). However, we emphasize that this upper limit is not physically constraining due to rapid water photodissociation in debris disk environments. The absence of observed water vapor in debris disks and unknown dissociation rates prevent meaningful interpretation of this value.

With these assumptions and the photodissociation effect unaccounted for, we infer that the cometary release of warm CO is a plausible mechanism, based on the warm CO-to-H$_2$O ratio in HD 131488 compared to the CO-to-H$_2$O ratio observed in solar system comets. We obtain a warm CO/H$_2$O mass ratio of $0.8$\% for HD 131488. For comparison, the CO/H$_2$O abundance in solar system comets ranges from $0.2$\% to over $10\%$ with a median of $(3\,\pm\,1)$\% \citep{MummaCharnley2011, Pinto+2022}. Our estimated $0.8$\% warm CO-to-H$_2$O ratio is consistent with the solar system comets' median value within $ 3\sigma$. Therefore, without accounting for the photodissociation effects, we infer that the cometary outgassing remains a viable mechanism to generate the trace amount of newly detected warm CO gas. We note that a full non-LTE calculation that takes into account photodissociation of water molecules around HD 131488 might be needed to place a constraining upper limit for the collision partner mass, and likely will increase the upper limit. 
\newline
\newline
\subsubsection{Atomic Gas as Collisional Partners}
We note the possibility of circumstellar atomic gas being a potential collisional partner to warm CO gas, but defer any quantitative calculations to future works. Circumstellar atomic gas of Ca and K has been observed in HD 131488 by \citet{Rebollido+18}. Although the distribution of Ca and K is not spatially resolved, these atomic gas species are inferred to extend radially to a few stellar radii and kinematically resolved at $4\,$km$\cdot s^{-1}$ based on radial velocities from high-resolution optical spectra. Similarly, recent JWST/MIRI observation of an archetypal debris disk, $\beta$ Pictoris, reveals spatially resolved atomic gas such as Argon, out to  $20$\,AU \citep{Worthen2024b}. In addition, JWST/MIRI observations reveal the presence of multiple atomic species, including atomic chlorine and sulfur, as well as ionized nickel, in the debris disk, HD 172555 \citep{Samland+2025}. While these collisional rates (e.g., CO-Ca, CO-K at $\sim1000$\,K temperatures) currently lack sufficient laboratory measurements, we note the possibility of CO being thermally excited by collisions with Ca and K atomic gas. We introduce this possibility at a superficial level and encourage future theoretical or laboratory studies to explore the role of atomic gas species in the evolution of molecular gas in debris disks. 

\section{Discussion}\label{sec5}
\subsection{Comparison with UV Fluoresced CO Gas in protoplanetary disks around Herbig Ae/Be Stars and T Tauri Stars}
The discovery of warm CO ro-vibrational emission in the debris disk HD 131488 links the inner disk region of debris disks to their predecessors, the inner disk regions of protoplanetary disks, along an evolutionary timeline.
Ro-vibrational emission of warm and hot CO molecules has been extensively studied to trace the structure, dynamics and time evolution of the innermost region of protoplanetary disks around Herbig Ae/Be stars and T-Tauri stars \citep[e.g., ][]{ najita2000, najita2003, najita2007, carr2001, carr2007, Brittain+03, Brittain+07, blake2004, salyk2009, salyk2011, bast2011, herczeg2011, vdp2015, banzatti2015, Banzatti+2022, hein2016}. The inferred inner gas radii from CO ro-vibrational lines have been found to mostly coincide with the inner dust radius in general \citep[see a review by][and reference therein]{Brittain+SSR} and also coincide with the presence of water lines in some T-Tauri disks \citep{Banzatti2017}. 
Recent near-infrared spectroscopic surveys further indicate that CO ro-vibrational emission is commonly detected around Herbig disks and T tauri disks \citep{Dickson-Vandervelde+2025}.

Compared to the high detection rate of fundamental ro-vibrational emission around Herbig Ae/Be stars and CTTS, with a typical gas temperature of $1000\,$K, this is the first time that CO ro-vibrational emission has been detected in a debris disk. In lieu of the search for CO in debris disks, 49 Ceti is another example where ro-vibrational emission has been detected albeit at a much lower temperature \citep{Worthen+2025}. In addition, spectroastrometric surveys of gas kinematics in Herbig star transition disks suggest that gas giants or brown dwarf perturbers are common \citep{Jensen+2024}. Therefore, the CO ro-vibrational emission and the gas kinematics could also be a useful probe for any hidden planetary perturber in debris disks. 

CO rovibrational emission lines in HD 131488 open a new door to probe the connection between the gas and dust structure and infer the potential existence of gas giants in the innermost region of this debris disk, leveraging tools developed by the protoplanetary disk community. The dust in the inner disk of HD 131488 inner region has been mostly studied via SED because the region interior to $\sim\,40\,$AU is inaccessible from ground-based coronagraphic images. Our model has shown that the CO ro-vibrational emission likely originates from a region that extends interior to $\sim\,10\,$AU given NIRSpec's pixel scale. While the search for planets is still ongoing in the outer disk of HD 131488 via direct imaging studies, the CO ro-vibrational lines give a unique opportunity to trace the inner disk dynamics via resolved line spectroscopy. Future spectrally-resolved, high-resolution M band study of HD 131488 can inform us of the rotational velocity of the warm CO population and its variability. 

In addition, HD 131488 shows no signs of accretion onto the young stars as expected. \citet{Melis+13} analyzes the optical spectrum of HD 131488 and finds neither noticeable emission features nor any obvious filling of the Balmer absorption lines in the spectrum. Additionally, based on H I Pfund lines at $4.6538\,\mu$m, which is a common accretion diagnostic in class I protoplanetary disks \citep{Salyk+2013}. In Fig. \ref{fig:COmodel}, we show a decomposition of the 3 components of our model. The solid gray lines show the stellar H I Pfund and Humphrey lines, in absorption, indicating a lack of accretion activity that is often seen in emission.

\subsection{Nature of the Gas: Primordial or Secondary?}
\begin{figure*}[ht!]
\epsscale{1.2}
    \plotone{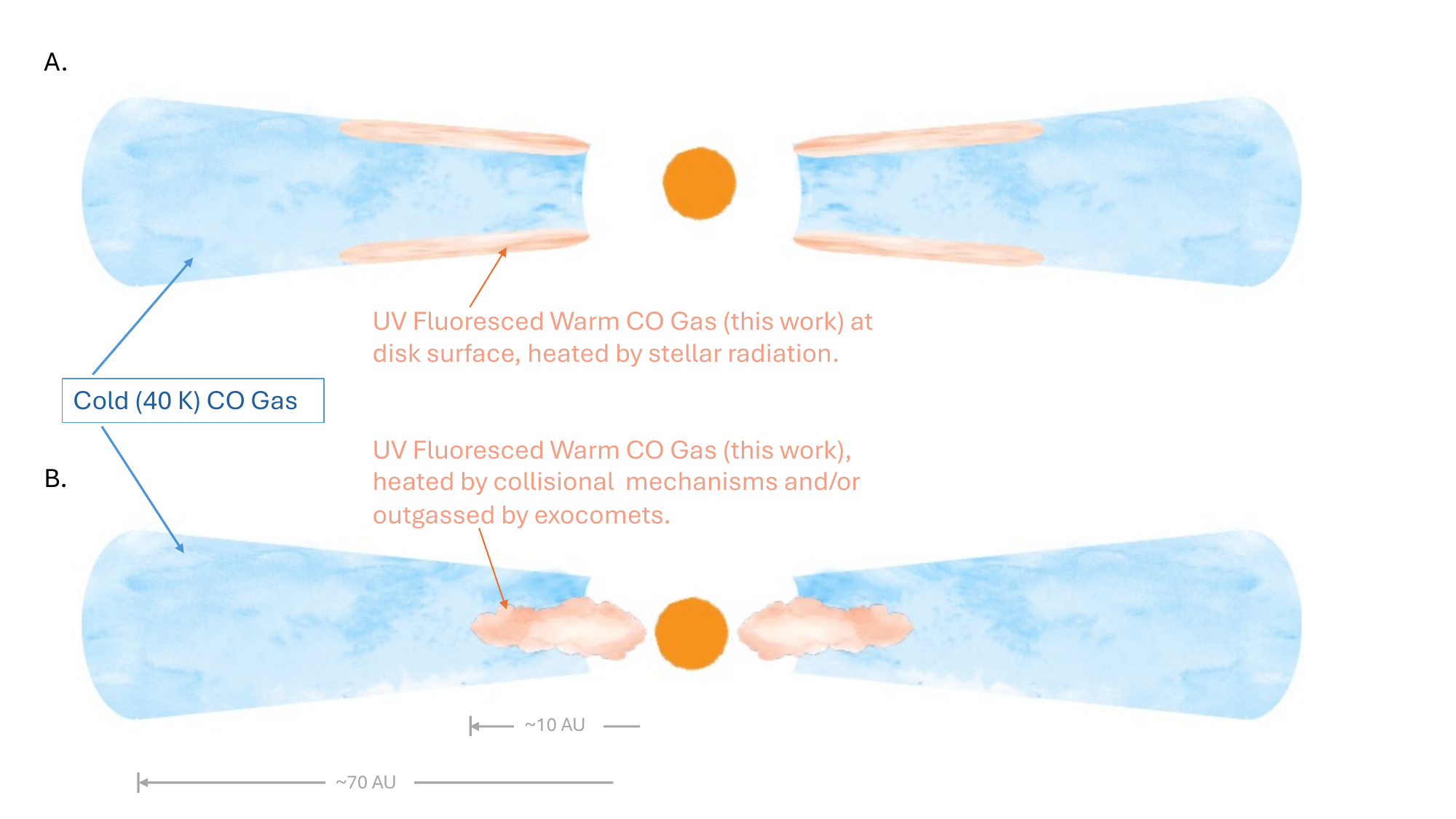}
    \caption{Illustrative figure for the warm CO detected in HD 131488 disk. \textbf{Panel A}: A plausible scenario is that the newly detected warm CO is spatially co-located with an optically thick cold gas disk, with the bulk of the gas residing beyond $\sim\,30$\,AU outwards. We are probing the skin layer of such a gas disk due to the optical depth effects of the CO fluorescence transitions. \textbf{Panel B}: An alternative scenario where the newly detected warm CO extends from $\sim10\,$AU to sub-AU ($0.2\,$AU) distance to the star. The warm gas would be potentially optically thin and generated from exocometary outgassing or collisional mechanisms. Given the emitting geometry of the CO gas derived from our best-fit models, this scenario appears more plausible.}
\label{fig:interpretation}
\end{figure*}
We discuss the possible origin of the warm CO gas in the context of the CI gas and cold ($\sim\,40\,$K) CO gas discovered with previous HST and ALMA observations \citep{Brennan+24, Moor+17}. The unshielded life time of CO in the interstellar radiation field is $\sim\,130$ years \citep{Heays2017}.
% and this timescale is even shorter with the effect of stellar radiation field. 
\citet{Brennan+24} performed simulations based on HST and ALMA measured gas column density in limits of high CO input and low CO input rates as a function of gas disk viscosity. Interestingly, with just the cold CO gas, the measured column density CI/CO (cold CO) is lower than predicted in a vertically well-mixed, second-generation gas release scenario \citep{Brennan+24}. Two hypotheses proposed by \citet{Brennan+24} are the removal of CI gas by reforming CO molecules via reactions with H$_2$ bearing molecules that could be either primordial or of exocometary origin (or via other unknown processes). \citet{Brennan+24} suggests the CI can be either layered or well-mixed with the cold CO gas. We build on those hypotheses to interpret our warm gas detection. Figure \ref{fig:interpretation} illustrates those two scenarios. Fig \ref{fig:interpretation} panel A illustrates one possible scenario in which the warm CO emission arises from the skin layer of the optically thick cold gas disk discovered with ALMA, which is predominantly located at radii beyond tens of AU ($\sim30$\,AU). In this scenario, the CI gas will be well-mixed with the warm CO at the skin layer of the disks, and the CO gas will primarily be heated by the stellar UV radiation and the interstellar radiation field. Future spatially resolved observations of \ion{C}{1} are needed to confirm this hypothesis. Fig \ref{fig:interpretation} B shows an alternative scenarios in which the newly detected warm CO extends from the vicinity of the star at sub-AU distance to $\sim10\,$AU. 
In this scenario, the warm CO should be optically thin and located at the mid-plane of the disk, and we are probing warm gas in non-LTE conditions that must be collisionally excited by a dark gas, which we expanded on in section \ref{sec:collision-partners}. 

We also explore the possibility of CO gas being of secondary origin by comparing the properties of the newly discovered UV-fluoresced warm CO population with the properties of gas seen in the solar system cometary coma. Compared to cometary gas which has a kinetic temperature range from tens of kelvin to a few hundred kelvins ($20$--$200\,$K) and a vibrational temperature equal to the diluted color-temperature of the Sun ($\sim\,5500\,$K), the warm CO in HD 131488 has a higher kinetic temperature of $125\,$--$\,450\,$K and a diluted color-temperature of its A0V host star $\sim 8800$ sub-AU regions and an average vibrational temperature of $5157\,$K, similar to the radiation field felt by the sun-grazing comets. However, it is worth noting that the gas survival time around an A-type star is approximately half that around a G-type dwarf, due to the combined effect of X-ray, near-UV, and far-UV radiation fields \citep{Nakatani+23}. Therefore, the photo-dissociation rate for molecules would be considerably faster in A0 star disks compared to G star disks. While CO$_2$ (with $\sim 4.3$--$4.5\,\mu$m ro-vibrational emission features) is a species commonly observed in solar system coma, we do not observe such features in our spectrum. 
Despite no detection of CO$_2$, our estimated upper limit for CO/H$_2$O ratio suggests that cometary outgassing remains a possible venue. The estimated upper limit $0.8$\% of CO/H$_2$O abundance ratio is close to the median values \citep[$3\,\pm\,1$
\%,][]{MummaCharnley2011, Pinto+2022} of CO/H$_2$O abundance ratio of salar system comets within $3\,\sigma$. Follow-up studies with ground-based high-resolution CO spectra are needed to spectrally resolve the emission lines to probe the radial velocity and inferred distance of the gas to test its association with the previously detected exocometary gas \citep[atomic species, Ca and K; ][]{Rebollido+18}. 

\subsection{Possible Production Timescale for the Warm CO Gas}
Since falling and evaporating bodies (also known as ``exocomets'') have been observed for the system \citep{Rebollido+20}, exocomets releasing gas is a possible production mechanism for the warm CO gas seen in HD 131488. We reference the simulations performed in \citet{Brennan+24} \citep[performed with \texttt{EXOGAS}; ][]{Marino+2020, Marino+2022}, where the authors show the high and low CO gas input rates at low temperatures. If we assume the warm gas is released from the same parent population planetesimals, we can use our column density and apply the results of their simulations to infer a warm gas production timescale. In the case of low gas disk viscosity, we find that our measured gas column density is consistent with a production timescale of $<0.1\,$Myr when both stellar and interstellar radiation fields are considered. In the case of high viscosity, the gas production timescale will be even shorter, which enables us to trace exocometary gas in action. If the gas is released from exocomets, simulation predicts asymmetric distribution of hot dust \citep{Bonsor+12Zodi}. Future spatially and/or spectrally resolved observations can help distinguish the contributions from exocomets from line velocity and spatial distribution of the gas. Furthermore, due to variability in comets' incident rates close to the star, it is likely that the future measured line fluxes could vary from the measured values in this work. If such variability is measured, then we can constrain the gas input rate for warm gas. In short, future time-domain follow-up observations are needed to understand the nature of the warm CO gas.

% \begin{itemize}
%     \item Talk about ALMA observation vs this work.
%     \item Effect of Slit Alignment on Gas Emission Features, the detected line flux is probably a lower limit
% \end{itemize}

\subsection{Warm gas in the Terrestrial Zone of Debris Disk as a Source of Metallicity in Exoplanet Atmospheres}
The discovery of warm gas in the terrestrial planet zone of a $10$ Myr debris disk raises the possibility of late-stage CO gas accretion as a source of planetary atmospheric metallicity. To date, a growing number of exoplanets have been discovered around young stars in the star-forming regions. Planets ranging in sizes from sub-Neptunes \citep[e.g. K2-33 b, AU Mic b, c; ][etc]{Rizzuto+2017, David+2016, Mann+2016, Plavchan+2020, Dai+2024} to super Jupiter \citep[e.g., PDS 70 b, c, $\beta$ Pictoris b, c; ][]{Lagrange+10, Lagrange+20, Nowak+20} have been found in gap-opening transition disks and debris disks. Recent observations of giant planets \citep[PDS 70c, Delorme AB b][]{Ringqvist2023, Zhou+2025} show that even after formation, giant planets are still capable of accreting gas at a few to tens of Myr. For small planets, \citet{Kral+20Nat} illustrates that the presence of a tenuous amount of gas could still be accreted onto small planets in $10$ Myr-year-old, gas-poor debris disk. Even for gas input rate as low as $10^{-5}$--$10^{-6}\,\text{M}_{\oplus} \cdot \text{Myr}^{-1}$, the gas accretion process could elevate the metallicity of a young protoplanet at 1/10 of Earth-mass from solar metallicity (C/O $\sim0.5$) to super solar metallicity (C/O$\sim0.6$--$0.8$). Our discovery of the presence of warm CO gas ($10^{-7} \,\text{M}_{\oplus}$) in the terrestrial zone possibly created in $0.1\,$Myr assuming a low viscosity gas disk, in a $10$ Myr-year-old debris disk, provide a good test bed for the simulation and motivates the search for planets at AU scale in the disk. 

\section{Conclusion}\label{sec6}
We present in this paper the detection of CO rovibrational emission from the circumstellar environment of the debris disk HD 131488. The lines observed in the JWST NIRSpec spectra seem to be a combination of both emission and absorption, and are probably the effect of two or more gas populations in the system, with different temperatures. Our UV fluorescence model gives a best fit of $1150\,K$  with an effective temperature of $450$, $332$, and $125\,$K for the warm CO gas kinetic temperature within $0.5$, $1$, and $10\,$AU and a $8800\,K$ vibrational temperature. We estimate a lower mass limit for CO of $1.25\times 10^{-7}\text{M}_{\oplus}$. The large discrepancy in CO's vibrational and rotational temperature indicates that CO is out of thermal equilibrium and is excited with UV fluorescence. The detection of warm CO raises the possibility of unseen molecules (H$_2$O, H$_2$, etc) as collisional partners to excite the warm gas. We conclude that the UV fluorescence of CO opens a new venue to search for warm gas in debris disks. 

\begin{acknowledgements}
CL thanks Matias Rodriguez and Winston Wu for helpful discussions. KW and CC acknowledge the support from the STScI Director's Research Fund  (DRF) and the NASA FINESST program. This work is supported by the National
Aeronautics and Space Administration under Grant No.
80NSSC22K1752 issued through the Mission Directorate.
AB acknowledges research support from the Irish Research Council under grant GOIPG/2022/1895. LM acknowledges funding by the European Union through the E-BEANS ERC project (grant number 100117693). Views and opinions expressed are however those of the author(s) only and do not necessarily reflect those of the European Union or the European Research Council Executive Agency. Neither the European Union nor the granting authority can be held responsible for them. This research has made use of the SIMBAD database, operated at CDS, Strasbourg, France. 

The JWST/NIRSpec data of HD 131488 presented in this article were obtained from the Mikulski Archive for Space Telescopes (MAST) at the Space Telescope Science Institute. The specific observations analyzed can be accessed via \dataset[doi: 10.17909/fjaj-7w30]{https://doi.org/10.17909/fjaj-7w30}.
\end{acknowledgements}

\software{\texttt{NumPy} \citep{numpy}, \texttt{Scipy} \citep{SciPy}, \texttt{Astropy} \citep{astropy:2013, astropy:2018, astropy:2022}, \texttt{emcee}\citep{emcee}}

\clearpage
\appendix
\section*{Appendix A}
We provide the MCMC corner plot for posterior distributions of the fitted parameters. For each parameter, the annotated text, its superscript, and subscript along the diagonal illustrate the best value, the 16th and 84th percentile as reported in Table \ref{tbl:best-fit}. The dashed lines also give visual cues to the range of 16th and 84th percentiles. Both the gas kinetic temperature $T_{\rm kin,0}$ and the $^{12}$CO column density, $\log_{10} (N)$ are sampled in log spaces. All other parameters are sampled in linear spaces.
\begin{figure*}[h!]
\epsscale{1.1}
    \plotone{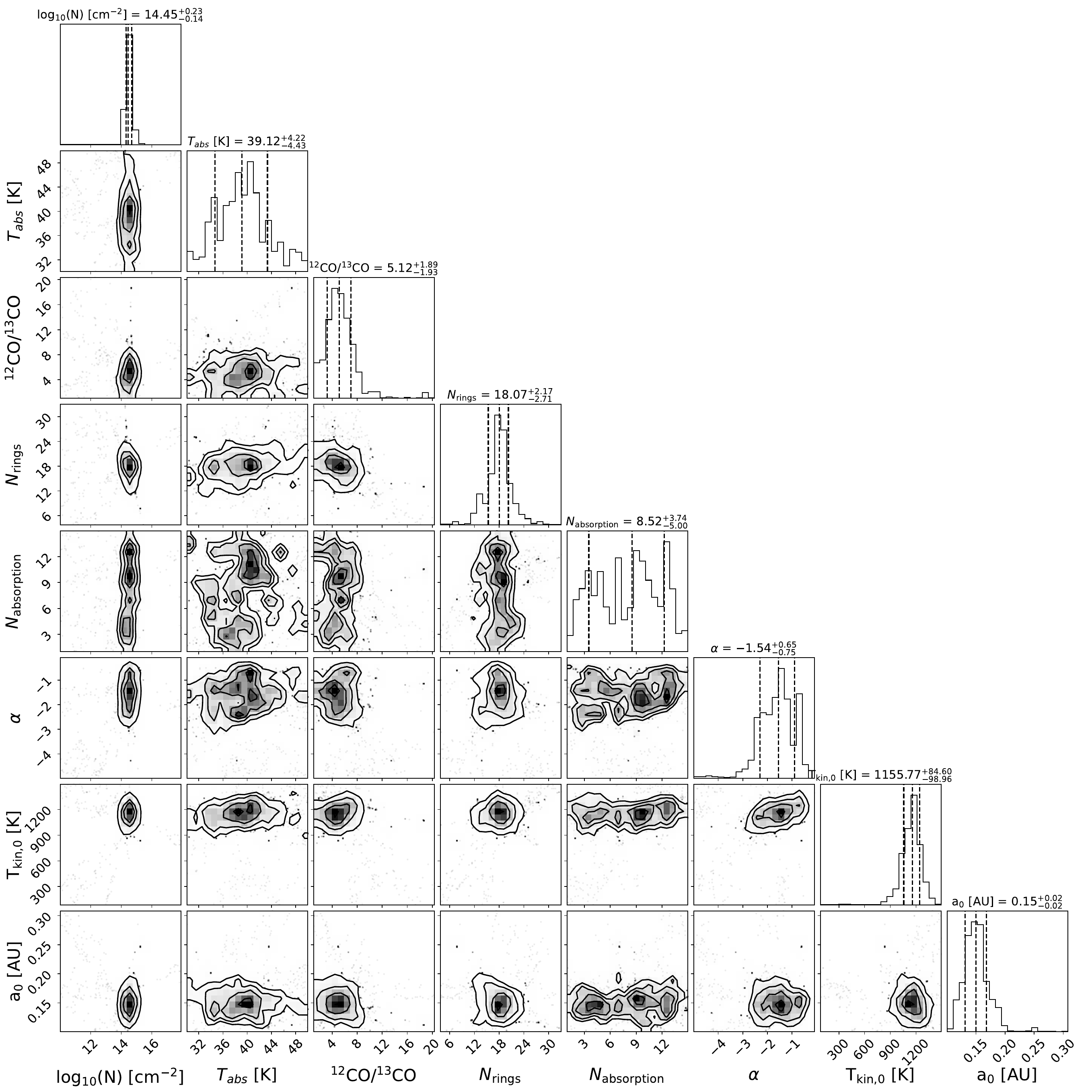}
    \caption{Corner Plot of Parameter Posterior Distribution for MCMC Sampling.}
    \label{fig:postDistribution}
\end{figure*}
\clearpage
\section*{Appendix B}
In Table \ref{tbl:line-flux}, we show diagnostics for line blending identification as described in section \ref{sec:line-flux}. We compare the rest frame wavelengths of various J transitions for the v(1-0) transition with lines observed at adjacent wavelengths. However, we quickly realized that essentially every line has some level of blending with higher vibrational transitions. Based on the common line blending issue, we avoided individual line-based approaches that are commonly used in ground-based, high-resolution spectroscopy works. 

In Figure \ref{fig:Model-bt}, we show a zoomed-in comparison between the model and the data. Our model provides a $\chi^2\sim1.01$ fit to the model. 

\begin{deluxetable*}{lcccccc}[th!]
\tablecaption{Observed and Rest Position, and Measured Flux for $^{12}$CO (1-0) Transitions} \label{tbl:line-flux}
\tablehead{
\colhead{Line ID} & \colhead{$\lambda_{obs}$} &\colhead{$\Tilde{\nu}_{obs}$}&  \colhead{$\lambda_{rest}$} & \colhead{$\Tilde{\nu}_{rest}$}& \colhead{F$_{obs}\pm\delta$F} & 
\colhead{Comments}\\
\colhead{} & \colhead{($\mu$m)} & \colhead{(cm$^{-1}$ )} & \colhead{($\mu$m)} & \colhead{(cm$^{-1}$)} &\colhead{($\mathrm{W}\cdot \mathrm{m}^{-2}$)}
& \colhead{}}
\startdata
R1(1-0) & $4.6494$ & $2150.8$ & $4.6493$ & $2150.9$ & $5.26\,\times 10^{-17}$ &\text{Blended with} $^{13}\text{CO}\,$ \text{R24(2-1)}\\
R4(1-0) & $4.6255$ & $2161.9$ & $4.6254$ & $2162.0$ & $5.26\,\times 10^{-17}$ & \text{Blended with} $^{13}\text{CO}\,$ \text{R28(2-1)}\\
R5(1-0) & $4.6175$ & $2165.7$ & $4.6177$ & $2165.6$ & $4.31\,\times 10^{-17}$ & \text{Slightly Blended with} $^{13}\text{CO}\,$ \text{R20(1-0)}\\
R9(1-0) & $4.5877$ & $2179.7$ & $4.5876$& $2179.8$& $4.45\,\times 10^{-17}$ &  \text{Slightly Blended with} $^{13}\text{CO}\,$ \text{R25(1-0)}\\
R10(1-0)$^{\clubsuit}$ & $4.5804$ & $2183.2$& $4.5804$& $2183.2$ & $4.40\,\times 10^{-17}$ & \nodata\\
R19(1-0) & $4.5187$& $2213.0$& $4.5195$& $2212.6$& $6.70\,\times 10^{-17}$ &\text{Blended with} $^{13}\text{CO}\,$ \text{R38(1-0)}\\ 
R44(1-0)$^{\clubsuit}$ &$4.3906$ & $2277.6$& $4.3901$& $2277.8$ &$7.27\,\times 10^{-17}$ & \nodata\\
\hline
P9(1-0) & $4.7449$ & $2107.5$ & $4.7451$ & $2107.4$ & $5.85\,\times 10^{-17}$ & \text{Slightly Blended with} $^{13}\text{CO}\,$ \text{R18(3-2)}\\
P10(1-0) & $4.7542$& $2103.4$& $4.7545$& $2103.3$& $5.90\,\times 10^{-17}$ & \text{Blended with} $^{13}\text{CO}\,$ \text{R1(1-0)}\\
P11(1-0)$^{\clubsuit}$ &  $4.7642$& $2099.0$ & $4.7640$ & $2099.1$ & $5.78\,\times 10^{-17}$& \nodata\\
P12(1-0)$^{\clubsuit}$ & $4.7734$& $2094.9$ & $4.7736$ & $2094.9$ &$5.74\,\times 10^{-17}$ & \nodata\\
P15(1-0)$^{\clubsuit}$ & $4.8033$ & $2081.9$ & $4.8031$ & $2082.0$ &$3.95\,\times 10^{-17}$ &\nodata\\ 
\hline
\enddata
\tablecomments{1. None of the lines listed in this table are blended with $^{12}\mathrm{CO}$ v(3-2), v(2-1), or v(1-0) transitions. \\ 2. $\clubsuit$ denotes the lines that don't blend with either $^{12}\mathrm{CO}$ or $^{13}\mathrm{CO}$ v(3-2), v(2-1), v(1-0) transitions and the line central wavelength are plotted in Fig \ref{fig:model_annotated}.}
\end{deluxetable*}
\begin{figure*}[h]
\epsscale{1.1}
    \plotone{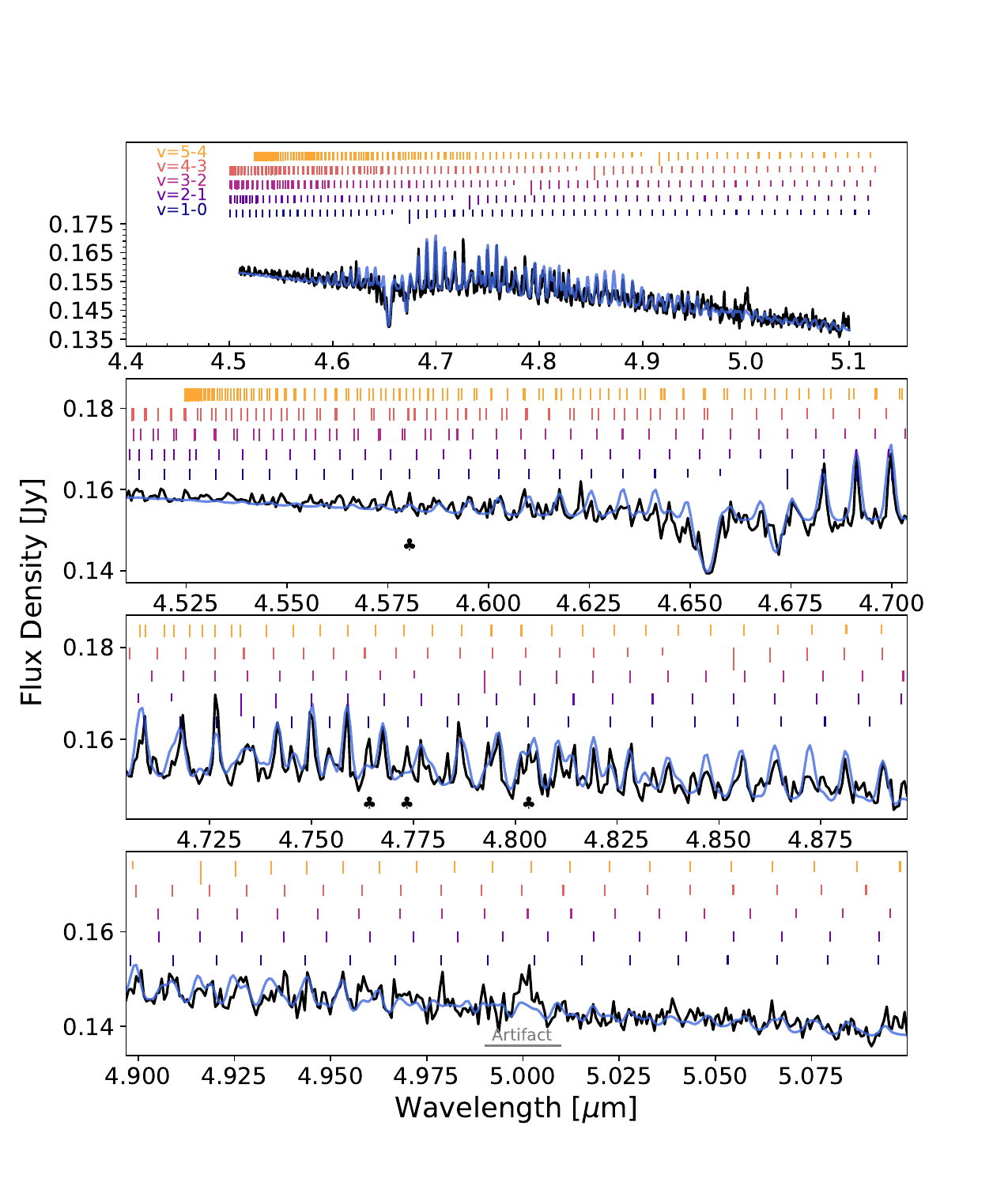}
    \caption{HD 131488 CO Model (in blue) plotted over NIRSpec data (in black). Similar to Fig.  \ref{fig:Model-bt}, but with CO ro-vibrational emission lines locations overlaid to illustrate the line-blending issue. The club $\clubsuit$ annotates locations of central wavelengths for lines in Table \ref{tbl:line-flux}. }
    \label{fig:model_annotated}
\end{figure*}

\begin{figure*}[th!]
    \centering
    \includegraphics[scale=0.7]{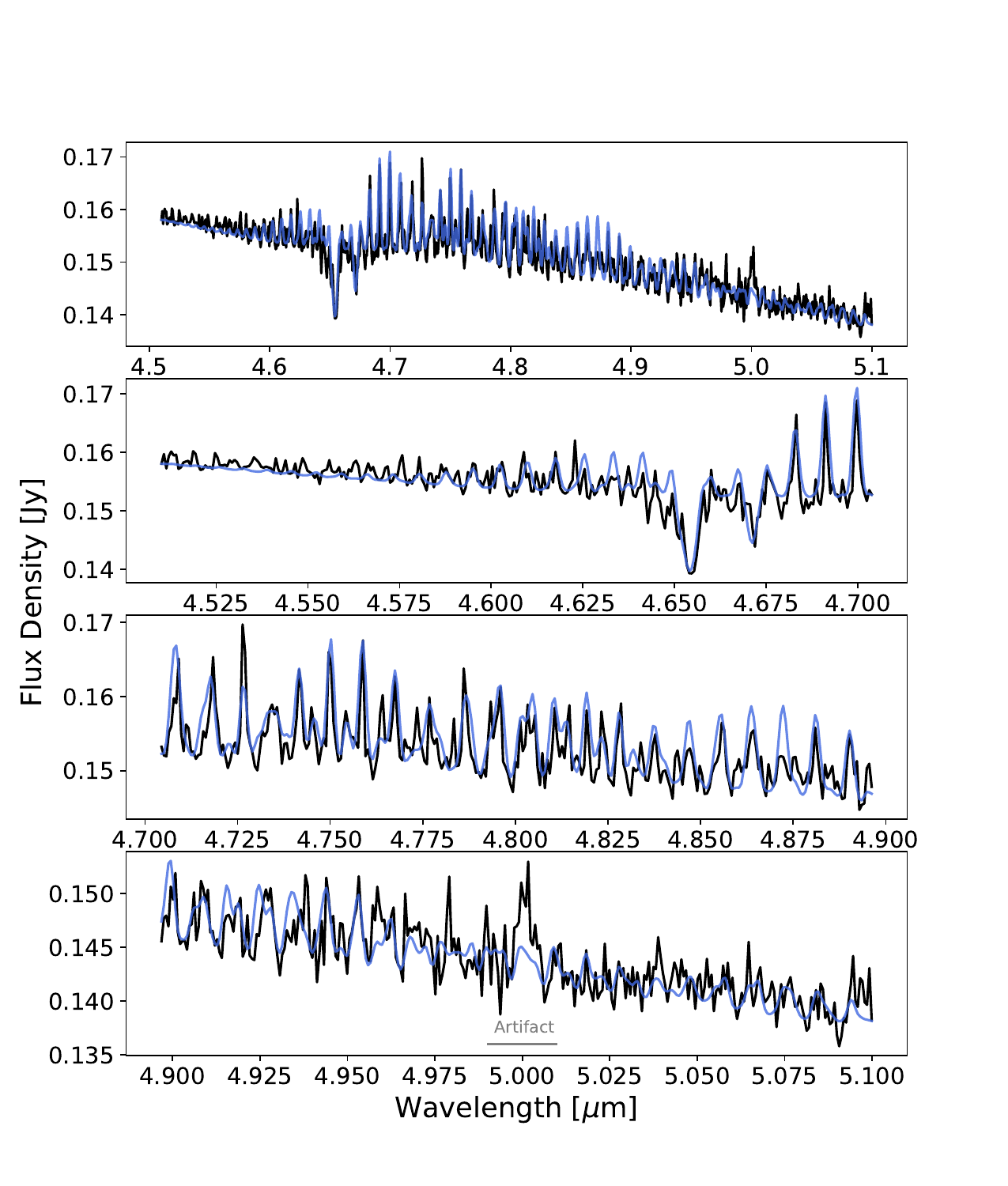}
    \caption{HD 131488 CO Model (in blue) plotted over NIRSpec data (in black). The top panel shows the same best-fit model as that of the top panel of Figure \ref{fig:COmodel}. The rest of the panels are zoomed-in regions of the spectral to clearly showcase the details of the agreement between model and data. It is worth noting that the emission line at $5\,\mu$m is under-fitted on purpose because the prominent line is likely a detector artifact.  These $5\,\mu$m lines exist in all spectra in JWST PID 2053 (PI: Rebollido) but at least one source's MIRI spectrum (GTO program, PI: Henning) at commensurate wavelength does not show such a feature.}
    \label{fig:Model-bt}
\end{figure*}
\clearpage

\appendix
\section*{Appendix C}
\begin{figure*}[th!]
    \centering
    \includegraphics[scale=0.6]{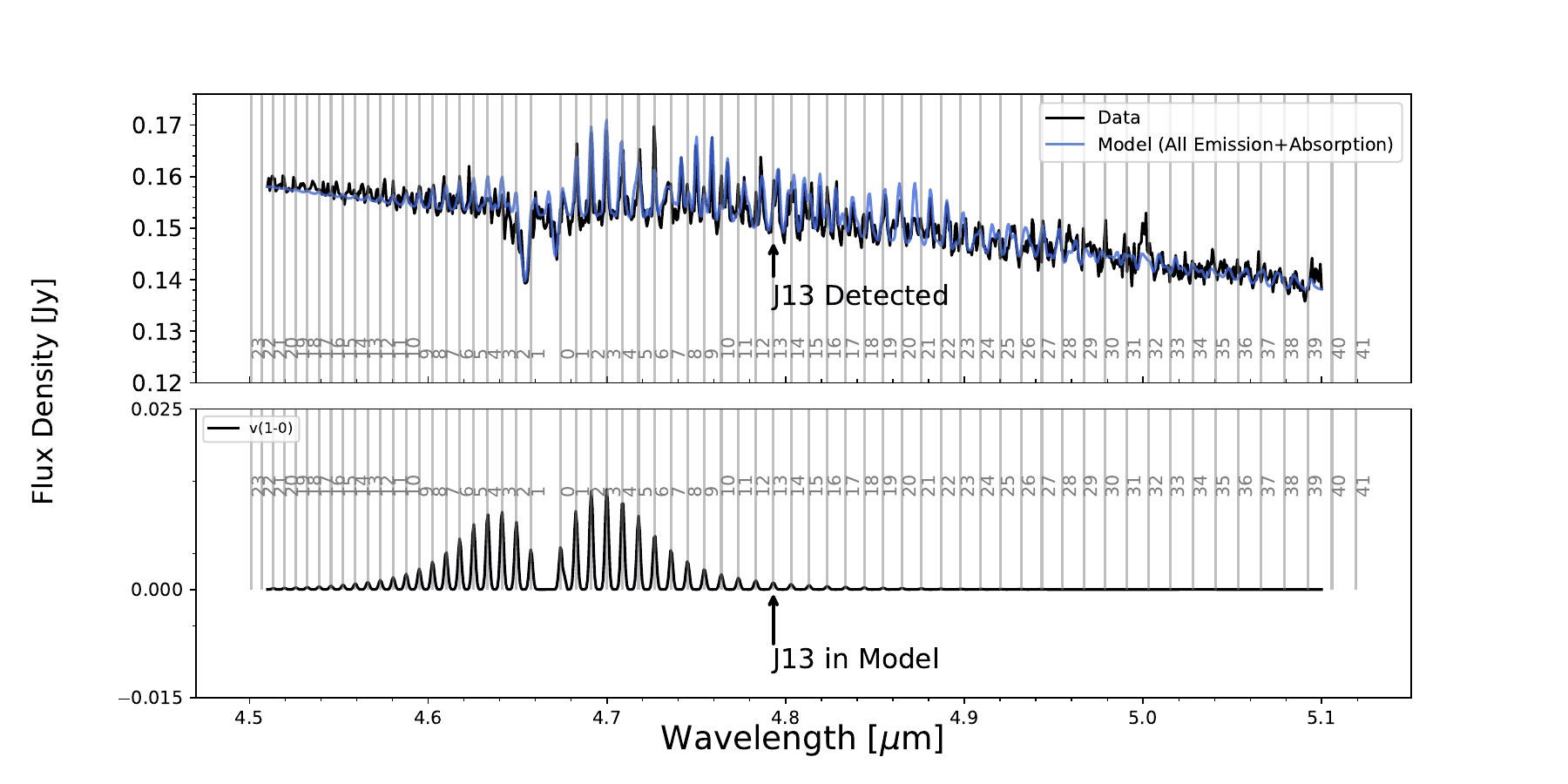}
    \caption{Highest J lines from the CO fundamental v(1-0) transition.}
    \label{fig:highJ}
\end{figure*}
In Figure \ref{fig:highJ}, we show the highest J transition detected for v(1-0). We then use this highest J line detected for estimating the critical density needed to collisionally excite CO with solely H$_2$ or H$_2$O molecules in section \ref{sec:collision-partners}. 

For a given molecular transition, the critical density is the density of colliders (H$_2$) at which the rate of collisional de-excitation equals the rate of spontaneous radiative decay. 

\begin{equation}
    n_{\rm crit} = \frac{A_{ul}}{C_{ul}},
\end{equation}
where $A_{ul}$ is the Einstein A coefficient for spontaneous emission (from upper level $u$ to lower level $l$). It's a fixed value for a specific transition. $C_{ul}$ is the collisional rate coefficient (in $cm^{3}\cdot\,s^{-1}$) for de-excitation by H$_2$. 
For a transition to be thermalized, the population of the energy levels is governed by the kinetic temperature of the gas (here we use rotational temperature to approximate the kinetic temperature). This happens when the gas density of H$_2$ is approximately $10 \times\,n_{\rm crit}$. This comes from a simple two-level system where the relative population of upper level compared to the perfect thermal equilibrium is given by,   
\begin{equation}
    \frac{n_u}{n_{u}^{therm}} = \frac{n_{H_2}}{n_{H_2}+n_{\rm crit}}.
\end{equation}
When $n_{H_2} = 10 \times\,n_{\rm crit}$, then the right hand side of the equation is $0.91$ where the upper level is now $91\%$ populated compared to the thermal equilibrium.

We can relate volume density in units of $cm^{-3}$ to column density by integrating along the line of sight such that $N_{CO} = n_{CO} \times L$ and $N_{H_2} = n_{H_2} \times L$. Therefore, 
\begin{equation}
    \frac{n_{H_2}}{n_{CO}} = \frac{N_{H_2}}{N_{CO}}. 
\end{equation}
To sub in $n_{H_2} = 10 \times\,n_{\rm crit, i}$, we obtain 
\begin{equation}
    \frac{N_{H_2}}{N_{CO}}\times n_{CO} \geq 10 \times n_{crit, i}. \label{method1}
\end{equation}
Rearranging the equation
\begin{eqnarray}
       & N_{H_2} \geq 10 \times n_{crit, i} \times \frac{N_{CO}}{n_{CO}} \nonumber\\
    & N_{H_2} \geq 10 \times n_{crit, i} \times L,\label{method2}
\end{eqnarray}
where $N_{H_2}$  is the inferred $H_2$ column density ($cm^{-2}$), $n_{crit, i}$ is the critical density for the CO(J=i→i-1) transition ($cm^{-3}$). $N_{CO}$ is the measured CO column density ($cm^{-2}$) and $n_{CO}$ is the volume density of CO ($cm^{-3}$). 

The challenge here is to realistically assess the warm CO volume density, as we only measured CO column density. We use two methods; (1) assume an abundance ratio between CO and H$_2$ and then follow equation \ref{method1}, and (2) use a characteristic scale length L from best-fit models and then calculate using equation \ref{method1}. For the former, a typical CO/H$_2$ abundance is $10^{-4}$, and for the latter, L is taken to be $\sim10$\,AU, the emitting radius of our best-fit model. We tabulate our results in Table \ref{tbl:collisions}. Note that the $N_{cp}$ is the lower limit to the column density, and therefore the ratio between N($^{12}$CO)/N$_{cp}$ is an upper limit to the unless N($^{12}$CO) is more massive than we measured.  

\begin{deluxetable}{ccccccc}[h]
\tablecaption{Collisional Partner Calculations}
\label{tbl:collisions}
\tablehead{
\colhead{Collisional} & \colhead{$A_{13,12}$}& \colhead{$C_{13,12}$@$300$\,K}&\colhead{$n_{\rm crit, 13\rightarrow12}$}& \colhead{$N_{\rm cp}$ assuming $10^{-4}$} & \colhead{$N_{\rm cp}$ using L= $10$AU} & \colhead{N(warm $^{12}$CO)/N$_{\rm cp}$ Range}\\
Partner & [$s^{-1}$]& [$cm^{-3}s^{-1}$]& [$cm^{-3}$]& [$cm^{-2}$] & [$cm^{-2}$]  & \nodata}
\startdata
H$_2$ & $1.7\times10^{-4}$& $3.8\times10^{-11}$ &$4.5\times 10^{6}$  & $1.7\times 10^{19}$ & $4.5 \times 10^{21}$& [$5\times 10^{-8}$,$10^{-5}$] \\
H$_2$O  & $2.5\times10^{-6}$& $8\times 10^{-11}$ & $3.1\times 10^{4}$  & $2.8\times 10^{18}$ &  $4.6 \times 10^{19}$& [1, 5]$\times10^{-4}$
\enddata 
\tablenotetext{a}{$N_{cp}$ stands for the estimated upper limit to the column density of collisional partner. }
\end{deluxetable}

\clearpage
\bibliographystyle{aasjournal}
\bibliography{main}
\end{document}